\documentclass[acmlarge,screen,nonacm]{acmart}
\usepackage{amsmath,amsfonts}
\usepackage{booktabs} 
\usepackage[10pt]{moresize}
\usepackage{graphicx}
\usepackage[english]{babel}
\usepackage{graphicx}
\usepackage{tcolorbox}
\usepackage{listings}
\usepackage{float}
\usepackage{multirow}
\usepackage[scaled]{helvet}
\usepackage{mathrsfs}
\usepackage[linesnumbered,ruled,lined,noend]{algorithm2e}
\usepackage{mathpartir}
\usepackage{mathtools}
\usepackage{dsfont} 
\usepackage{stmaryrd}
\usepackage{url}
\usepackage{textcomp} 
\usepackage{bbm}
\usepackage{verbdef}
\usepackage{xspace}
\usepackage{verbatim}
\usepackage{lipsum}
\usepackage{wrapfig}

\usepackage{enumitem}
\usepackage{pifont}
\usepackage{varwidth}
\usepackage{xpatch}
\usepackage{multirow}
\usepackage{xcolor}
\usepackage[export]{adjustbox}
\usepackage{caption}
\usepackage{subcaption}

\usepackage[varqu]{zi4}

\usepackage{flushend}

\usepackage{tikz}
\usetikzlibrary{calc,positioning,math,tikzmark}
\usetikzmarklibrary{listings}

\newtheorem{definition}{Definition}

\def\BibTeX{{\rm B\kern-.05em{\sc i\kern-.025em b}\kern-.08em
    T\kern-.1667em\lower.7ex\hbox{E}\kern-.125emX}}

\setlist[itemize]{topsep=0pt,itemsep=-1ex,partopsep=1ex,parsep=1ex,itemindent=0pt,leftmargin=8pt}

\definecolor[named]{ACMBlue}{cmyk}{1,0.1,0,0.1}
\definecolor[named]{ACMYellow}{cmyk}{0,0.16,1,0}
\definecolor[named]{ACMOrange}{cmyk}{0,0.42,1,0.01}
\definecolor[named]{ACMRed}{cmyk}{0,0.90,0.86,0}
\definecolor[named]{ACMLightBlue}{cmyk}{0.49,0.01,0,0}
\definecolor[named]{ACMGreen}{cmyk}{0.20,0,1,0.19}
\definecolor[named]{ACMPurple}{cmyk}{0.55,1,0,0.15}
\definecolor[named]{ACMDarkBlue}{cmyk}{1,0.58,0,0.21}

\makeatletter
\def\arcr{\@arraycr}
\makeatother

\makeatletter
\newcommand{\mtmathitem}{%
\xpatchcmd{\item}{\@inmatherr\item}{\relax\ifmmode$\fi}{}{\errmessage{Patching of \noexpand\item failed}}
\xapptocmd{\@item}{$}{}{\errmessage{appending to \noexpand\@item failed}}}
\makeatother

\definecolor{shadecolor}{gray}{1.00}
\definecolor{ddarkgray}{gray}{0.75}
\definecolor{darkgray}{gray}{0.30}
\definecolor{light-gray}{gray}{0.87}

\newcommand{\ie}{\emph{i.e.}\xspace}

\newcommand{\eg}{\emph{e.g.}\xspace}

\newcommand{\vs}{\emph{vs.}\xspace}

\newcommand{\cf}{\textit{cf.}\xspace}
\newcommand{\wrt}{\emph{wrt.}\xspace}

\newcommand{\True}{\mathsf{True}}

\newcommand{\set}[1]{\left\{{#1}\right\}}

\newcommand{\letz}{\mathsf{{\bf let}}\xspace{}}

\newcommand{\eqdef}{\triangleq}

\newcommand{\tname}[1]{\textsc{#1}\xspace}
\newcommand{\pfix}{\tname{PFix}}
\newcommand{\afix}{\tname{AFix}}

\newcommand{\hfix}{\tname{HFix}}
\newcommand{\grail}{\tname{Grail}}
\newcommand{\footpatch}{\tname{FootPatch}}

\newcommand{\inferr}{\tname{Infer}}
\newcommand{\racerd}{\tname{RacerD}}
\newcommand{\tool}{\tname{Hippodrome}}

\newcommand{\spath}{\pi}
\newcommand{\trace}{\tau}
\newcommand{\rrd}{\mathbf{rd}}
\newcommand{\rwr}{\mathbf{wr}}

\newcommand{\csummary}[8]{\langle #1,#2,#3,#5,#6,#8 \rangle}

\newcommand{\bugsad}{\mathcal{B}} 
\newcommand{\bug}{b} 
\newcommand{\bugset}{B} 
\newcommand{\patch}{p}
\newcommand{\act}{\mathit{act}}
\newcommand{\fpatch}{\widehat{\patch}}
\newcommand{\fact}{\widehat{\act}}
\newcommand{\patchset}{P}
\newcommand{\stmt}{s}
\newcommand{\snapshot}{a}
\newcommand{\algo}[1]{\textsc{#1}}

\newcommand{\rwbug}{Read/Write}

\newcommand{\acomment}[1]{{\color{gray}#1}}

\definecolor{pblue}{rgb}{0.13,0.13,1}
\definecolor{pgreen}{rgb}{0,0.5,0}
\definecolor{pred}{rgb}{0.9,0,0}
\definecolor{pgrey}{rgb}{0.46,0.45,0.48}

\definecolor{ckeyword}{HTML}{7F0055}
\definecolor{ccomment}{HTML}{3F7F5F}
\definecolor{cnumber}{HTML}{2A0099}
\definecolor{darkgreen}{HTML}{008000}

\lstdefinelanguage{Java}{
  keywords={new, return, public, this, if, else, try, then, private,
    push, pop, unlock, while, throw,
    synchronized, final, class, case, switch, new, for, protected, extends, implements, import,
    volatile, null, static},
  ndkeywords={bool, int, void, double, float},
  moredelim=[il][\textcolor{ACMOrange}]{$$},
  string=[b]",
  showspaces=false,
  showtabs=false,
  breaklines=true,
  showstringspaces=false,
  breakatwhitespace=true,
  lineskip=-0.0pt,
  morecomment=[l]{//}, 
  morecomment=[s]{/*}{*/}, 
  basewidth={0.54em, 0.4em},%
  basicstyle=\footnotesize\ttfamily,
  keywordstyle={\color{ACMPurple}\ttfamily\bfseries},
  ndkeywordstyle={\color{ACMDarkBlue}\ttfamily\bfseries},
  commentstyle={\color{ccomment}\itshape},
  stringstyle=\color{darkgreen},
  xleftmargin=1em,
  moredelim=[is][\textcolor{pgrey}]{\%\%}{\%\%}
}

\lstdefinestyle{numbering}{
  numbers=left,
  numberstyle={\scriptsize\color{cnumber}\sf},
  xleftmargin=2em
}

\newcommand{\jcode}[1]{\lstinline[language=Java,basicstyle=\small\ttfamily]{#1}}
\newcommand{\code}[1]{\jcode{#1}}

\def\unowned{\mathrm{Unowned}}
\newcommand{\ownedif}[1]{\mathrm{OwnedIf}(#1)}

\newcommand{\mcode}[1]{\text{\lstinline[language=Java,basicstyle=\small\ttfamily]{#1}}}
\newcommand{\mmcode}[1]{\text{\lstinline[language=Java,basicstyle=\footnotesize\ttfamily]{#1}}}

\newcommand{\race}{\mathit{race}}
\newcommand{\wrace}{\mathit{unprotected\_write}}
\newcommand{\cls}{\it{cls}}
\newcommand{\acc}{\it{acc}}

\newcommand{\idn}[1]{\mathit{#1}}

\hypersetup{colorlinks,
  linkcolor=ACMDarkBlue,
  citecolor=ACMPurple,
  urlcolor=ACMDarkBlue,
  filecolor=ACMDarkBlue}
  
\definecolor{dkgreen}{rgb}{0,0.6,0}
\definecolor{gray}{rgb}{0.5,0.5,0.5}
\definecolor{mauve}{rgb}{0.58,0,0.82}
\definecolor{backcolour}{rgb}{0.95,0.95,0.92}

\title{\tool: Data Race Repair using Static Analysis Summaries
}

\author{Andreea Costea}
\affiliation{
	 \institution{National University of Singapore}
	 \city{Singapore}
	 \country{Singapore}
	}
\email{andreeac@comp.nus.edu.sg}
\author{Abhishek Tiwari}
\affiliation{
	\institution{National University of Singapore}
	\city{Singapore}
	\country{Singapore}
}
\email{tiwari@comp.nus.edu.sg}
\author{Sigmund Chianasta}
\affiliation{
	\institution{National University of Singapore}
	\city{Singapore}
	\country{Singapore}
}
\email{sigmund@u.nus.edu}
\author{Kishore R}
\affiliation{
	\institution{National University of Singapore}
	\city{Singapore}
	\country{Singapore}
}
\email{kishore_r@u.nus.edu}
\author{Abhik Roychoudhury}
\affiliation{
	\institution{National University of Singapore}
	\city{Singapore}
	\country{Singapore}
}
\email{abhik@comp.nus.edu.sg}
\author{Ilya Sergey}
\affiliation{
	\institution{Yale-NUS College and National University of Singapore}
	\city{Singapore}
	\country{Singapore}
}
\email{ilya.sergey@yale-nus.edu.sg}

\begin{abstract}
Implementing bug-free concurrent programs is a
challenging task in modern software development.
State-of-the-art static analyses find hundreds of concurrency bugs in
production code, scaling to large codebases.
Yet, fixing these bugs in constantly changing codebases represents a
daunting effort for programmers, particularly because a fix in the
concurrent code can introduce other bugs in a subtle way.

In this work, we show how to harness compositional static analysis for concurrency bug detection, to
enable a new Automated Program Repair (APR) technique for data races in large concurrent Java codebases.
The key innovation of our work is an algorithm that translates
procedure summaries inferred by the analysis tool for the purpose of bug reporting, into small local patches that fix concurrency bugs (without introducing new ones). This synergy makes it possible to extend the virtues of compositional static concurrency analysis to APR, making our approach
\emph{effective} (it can detect and fix many more bugs than existing
tools for data race repair), \emph{scalable} (it takes seconds to
analyse and suggest fixes for sizeable codebases), and
\emph{usable} (generally, it does not require annotations from the users and can perform \emph{continuous} automated repair).
Our study conducted on popular open-source projects has confirmed that our tool automatically produces concurrency fixes similar to those proposed by the developers in the past.

\end{abstract}

\begin{document}
	\maketitle
	
\vspace{-5pt}

\section{Introduction}
\label{sec:introduction}

It is well acknowledged 
that implementing both correct and efficient concurrent programs is 
difficult~\cite{Herlihy-Shavit:08}.
While programmers have a robust understanding of sequential programs,
their understanding of concurrently interacting processes is often incomplete,
which may lead to  subtle bugs.
Once introduced, these bugs are hard to identify due to the inherently
non-deterministic nature of concurrent executions.
In other words, these issues can
only be detected under selective thread schedulings which are challenging to reproduce during debugging.
A number of tools have been
introduced to (semi-) automatically detect concurrency bugs in real-world programs, thus facilitating their discovery and
reproducibility~\cite{BlackshearGOS18,GorogiannisOS19,NaikAW06,Naik-al:ICSE09,
  FlanaganG05,RacerX,ClangTSA,TSAN, Flanagan-Freund:PLDI09,SamakRJ15}.

Successfully identifying a concurrency bug, via a tool or by manually
examining the code, does not, however, necessarily mean that a correct
fix for it is immediately apparent to the developer.
Even worse, by 
eliminating a data race between memory accesses to the same location,
it is not uncommon to introduce violation of other crucial properties, such as introducing a 
potential deadlock between threads.

Automated program repair (APR)~\cite{GouesPR19} is an emerging technology paradigm for automatically fixing bugs via search, semantic reasoning and learning. 
In the area of concurrent programs, APR has been employed to address the problem of maintaining multi-threaded
software and reducing the cost of migrating sequential code to a concurrent execution model~\cite{JinSZLL11,Jin0D12,Liu-al:FSE14,LinWLSZW18,ConTest,ConTest2,HuangZ12a,LiuZ12}.
Some tools eschew the issue of detecting bugs, focusing exclusively on
repair techniques and assuming that the bug descriptions are already
available, \eg, from a bug tracker or in a form of a dynamic execution
trace~\cite{JinSZLL11,Jin0D12,Liu-al:FSE14,LiuCL16,LiuZ12}. 
Other tools rely on their own approaches for bug detection, based on
run-time analysis or bounded model checking, and follow-up the
detection with a specific way of generating
patches~\cite{LinWLSZW18,KhoshnoodKW15,ConTest,ConTest2,HuangZ12a}.
However, these tools' reliance on dynamic analyses or bounded model checking for bug
detection poses significant challenges to their
adoption at large. 
First, it makes it problematic to integrate
 APR  into everyday development process with low
friction, as the developers are required to provide inputs for dynamic
executions or structure their tests accordingly, to enable bug
detection in the first place. 
Second, the lack of modularity prevents them from providing incremental feedback to the
programmers in the style of continuous integration~(CI). 

In this work, we address these challenges by describing the first
approach to perform APR for concurrency using 
\emph{static program analysis}.
Studies on static analysis usage at Facebook \cite{DistefanoFLO19}
and Google \cite{Sadowski-al:CACM18} show that a developer is roughly 70\%
likely to fix a bug if presented with the issue at compile time, as compared to 0\% to 21\%
fix attempts when bug reports are provided for checked-in code. 
The use of static analysis for concurrency bugs detection in industry is motivated by its success 
in catching bugs at scale
while minimising \emph{friction} (\ie, adoption effort) and providing
\emph{high signal} (\ie, useful bug reports)~\cite{AyewahHMPP08,BesseyBCCFHHKME10,AndyChou14}.
The high signal (\ie, low false positives rate) of such analyses is thanks to
the design choice to focus on the most common, \emph{coarse-grained},
concurrency (scoped locks and Java's \code{synchronized} blocks). Restricting
the class of analysed concurrent programs this way makes accurate static
detection of data races tractable, leading to accurate bug reports on industrial
code.
Scalability is achieved by 
making the analysis
\emph{compositional}~\cite{BlackshearGOS18}:
individual program components (\eg,
classes and methods) are analysed in a bottom-up, divide-and-conquer
fashion and abstracted as
\emph{summaries}. Summaries contain relevant information about the
underlying code, which does not have to be re-analysed
again. 
Furthermore, this design favours modularity: the analysis can be
executed incrementally, providing nearly instant feedback on recent
code changes to the developers.

Our novel approach to APR for concurrent programs builds on a compositional
static analysis for data race detection.
As the result, our repair tool captures and effectively navigates the fix space
for data-race repair.
The two main technical challenges we overcome in this work are (a) enhancing a
state-of-the-art static analysis for concurrency to collect sufficient
information to produce correct concurrent patches efficiently and at scale, and
(b)~devising a family of algorithms that construct concurrency fixes from the
code summaries produced by the augmented analysis.
Building on \racerd, an industry-grade static
concurrency analyser by Facebook~\cite{BlackshearGOS18}, our tool, called
\tool,\footnote{Historically, Hippodrome was a place where chariot races
were decided.} implements an automatic repair procedure for \emph{data races} in
concurrent Java programs. 

The static approach endows our technique with several advantages.
First, it often requires \emph{no additional input} from the users besides
the program itself---enabling smooth
integration with the CI workflow.
Second, our approach enjoys the underlying analysis modularity,
producing fixes in a matter of seconds, thus, \emph{scaling to large
  real-world codebases}, and allowing for incremental code processing.
Finally, the soundness guarantees of the analysis (that hold under
certain side conditions), extend to the produced patches: by
re-running the analysis on the repaired code we ensure the
\emph{correctness of the suggested fixes}: the fixed program satisfies both data-race freedom and deadlock freedom.

To summarise, this work makes the following contributions:

\begin{itemize}

\item \emph{Concurrency repair from analysis specifications.}
  We present a series of algorithms that take the summaries produced by
   a static analysis for concurrent code
  and turns them into suggestions
  for possible fixes, thus delivering the first modular program repair
  procedure for data races based on a static analysis for concurrency.
  The design of our patch-generating algorithms addresses a number of
  pragmatic concerns, minimising the amount of synchronisation to be
  added for eliminating races, while avoiding deadlocks
  (\autoref{sec:repair}).

\item \emph{Data race repair tool.}
  We make the implementation of our approach in a tool called \tool
  publicly available~\cite{racerdfix} for experiments and extensions.

\item \emph{Extensive evaluation.}
  We evaluate \tool on a number of 
  micro-benchmarks used by related tools for concurrent program
  repair~\cite{LinWLSZW18}, as well as on two popular large open-source
  projects (\autoref{sec:eval}).

\end{itemize}

\section{Motivation and Overview}
\label{sec:overview}

\begin{figure}[t]
\begin{subfigure}{0.6\textwidth}
\begin{lstlisting}[style=numbering]
public void run() {
  if (getError() == null) {
    try {
//   synchronized (this) { -- Second (correct) fix (1/2)
      if (read) {
         nBytes = getSocket().read(buffers, ...);
         updateLastRead();
      } else {
         nBytes = getSocket().write(buffers, ...);
         updateLastWrite();
      }
//   }                           -- Second (correct) fix (2/2)
// More code
}
\end{lstlisting}
\end{subfigure}
\hfill
\begin{subfigure}{0.39\textwidth}
\begin{lstlisting}
// First (faulty) fix and its commit message
- public void run() {
+ public synchronized void run() {
\end{lstlisting}
\begin{center}
\begin{tcolorbox}[width=\textwidth,colback={white},
title={\textbf{Add sync when processing asynchronous operation in NIO.}},
colbacktitle=white,
coltitle=black,
boxrule=0.5pt]    
{
{{
\noindent
%
The NIO poller seems to create some unwanted concurrency, causing rare
CI test failures [...] It doesn't seem right to me that there is
concurrency here, but it's not hard to add a sync.}}
}
\end{tcolorbox} 
\end{center}
\end{subfigure}
\caption{A data race in Apache Tomcat and its fixes. }
\label{fig:tomcat}
\end{figure}

In this section, we motivate and outline our approach for concurrency repair based
on a compositional static analysis. 

\subsection{Concurrency Bugs in the Wild}
\label{sec:tomcat}

To set the stage for our motivational case study of using a static
analysis to detect real 
concurrency issues, consider
the example in \autoref{fig:tomcat}.
The figure tells a curious story of a data race in the codebase of the
Apache Tomcat project.
The bug in this code snippet originates from the corruption of
buffers; two threads may simultaneously read and write into buffers
from the socket.
The bottom of the figure shows the faulty
commit
with a non-fix, which has been first attempted in order to remedy the
issue.
As the enclosed commit message makes evident,
this bug was rarely
observed, and the developer assigned to fix it was unable to
understand the exact root cause of the problem.
To make the situation worse, the developer simply made the whole
method \code{run} synchronised. 
This ``fix'' removed all concurrency whatsoever as \code{run} is the
entry method for threads, which all would be forced to run
sequentially now.
A closer look at the commit message reveals the lack of awareness
regarding the \code{synchronized} primitives.
A correct fix by the same developer is shown in the comments.

While this is just one example of a concurrency issue with a wrong
human-proposed solution and a simple correct one, there have been
multiple instances in the past where an incorrect fix caused other
severe problems, \eg, by introducing deadlocks. 
Thus, there is a need for reliable automated fixes that can guide the
developers towards the correct fix. For example, in this bug,
an automated fix suggestion (similar to the correct commit) could have
helped to avoid the wrong fix.

What gives us hope that there exists a way to engineer an APR
procedure for a large class of real-world concurrency bugs is the
following observation.
The valid fix in \autoref{fig:tomcat} is pleasantly \emph{simple}: it
just wraps the subject of the data race into Java's
\code{synchronized} block, thus introducing a necessary mutual
exclusion region in the otherwise parallelisable implementation. In
fact, it is so simple that we might hope to discover it
automatically.

\subsection{Dynamic Analysis-based Bug Detection}
To demonstrate why automatically characterising concurrency bugs in a
repair-friendly way is technically non-trivial, consider the example in
\autoref{fig:datarace}, taken from a suite of buggy
programs
\cite{pecan} used to showcase tools for concurrent
bug detection and APR~\cite{HuangMR14,LinWLSZW18}.
Only relevant code parts involved in a data race are shown.
In this example, threads may withdraw or deposit in multiple
accounts. However, there is a data race in both methods (between
line~\ref{line:1} and~\ref{line:2}, line~\ref{line:3}
and~\ref{line:4}). While one thread may read the
balance (line~\ref{line:3}), another may modify it
(line~\ref{line:4}). The data race is present in both methods, but
the testing thread only checks for the \emph{deposit} method
(line~\ref{line:5}). 
Thus, tools relying on test-based bug detection would miss the race in \code{withdraw}. 
While test harness could be improved, concurrency bugs in large-scale projects 
often go undiscovered due to the scheduling problem's intractability.  
For instance, \pfix~\cite{LinWLSZW18}, a recent tool for concurrent
APR, uses Java PathFinder (JPF)~\cite{VisserPK04} for bug
localisation, but unfortunately, JPF fails to detect the bug in Tomcat
(\autoref{fig:tomcat}) 
as it does not scale well to large programs.

\subsection{Overview of our Static Approach}
\label{sec:comp-conc-analys}
We next list the principles that make static analysis suitable 
for fault localisation in large
code. %
First, the class of bugs should be well-defined
on the premise that it is natural to give up on detecting \emph{all kinds} of 
bugs
and focus solely on one, \eg, \emph{data races} as per the current work.
Second, reducing the number of false alarms is crucial even when this  
entails giving up on the soundness 
(or framing soundness \wrt a set of assumptions),
 a compromise developers accept. 
Third, to scale up, the analysis should be compositional,
where the analysed units of code 
are ascribed \emph{summaries}, 
 allowing to re-analyse
modified code parts incrementally.
For data race detection, it suffices to summarise the
\emph{memory accesses} and the  \emph{corresponding locks} held at the access sites---thus, achieving
compositionality and, hence, scalability.

\begin{figure}[t]
\centering
\hspace{-2cm}
\begin{subfigure}{0.65\textwidth}
\begin{lstlisting}[style=numbering,firstnumber=1,xleftmargin=.25\textwidth]
 public class CustomerInfo {  
   private Account[] accounts;
\end{lstlisting}
\begin{lstlisting}[xleftmargin=.25\textwidth]
   // More fields and methods
\end{lstlisting}
\begin{lstlisting}[style=numbering,firstnumber=21,xleftmargin=.25\textwidth]
   public void withdraw(int accountNumber, int amount) {
 *\label{line:1}*    int temp = accounts[accountNumber].getBalance();
     temp = temp - amount;
 *\label{line:2}*    accounts[accountNumber].setBalance(temp);
   }

   public void deposit(int accountNumber, int amount) {
 *\label{line:3}*    int temp = accounts[accountNumber].getBalance();
     temp = temp + amount;
 *\label{line:4}*    accounts[accountNumber].setBalance(temp);
   }
 }

 public void run() {*\label{line:5}*  ci.deposit(1, 50);  } //Testing Thread

\end{lstlisting}
\end{subfigure}
\hspace{-0.5cm}
\begin{subfigure}{0.45\textwidth}
\begin{lstlisting}[style=numbering,firstnumber=1,xleftmargin=.25\textwidth]
 public class Account {
	
   private int balance;
		
  public int getBalance() {
 *\label{line:a1}* 	return balance;
  }
		
  public void setBalance(int balance) {
 *\label{line:a2}* 	this.balance = balance;
  }		
 }
\end{lstlisting}
\end{subfigure}
\caption{A data race example from a standard suite~\cite{LinWLSZW18}.}
\label{fig:datarace}
\end{figure}

Unfortunately, the textual bug reports provided by static analysis are not
immediately amenable for program repair as they contain too little contextual
information.
To remedy this, we process the internal
summaries of the analysis, extracting the necessary information 
from them, while solving two key challenges.
The first challenge is the \textit{selection of a suitable lock}.
A data race is avoided by protecting the affected memory access
via the same lock object.  
However, choosing this lock
statically is not trivial: there might be different locks to choose from, or none at all, as is the case in \autoref{fig:datarace}. 
Besides, new locks should not introduce deadlocks.
Second, handling the \textit{scope of the synchronisation} 
is equally important. 
Multiple locks may be combined into a single one, producing a concise patch. At the
same time, excessive locking inevitably hurts parallelism.

\begin{figure}[t]
	\centering
	\includegraphics[width=0.8\textwidth]{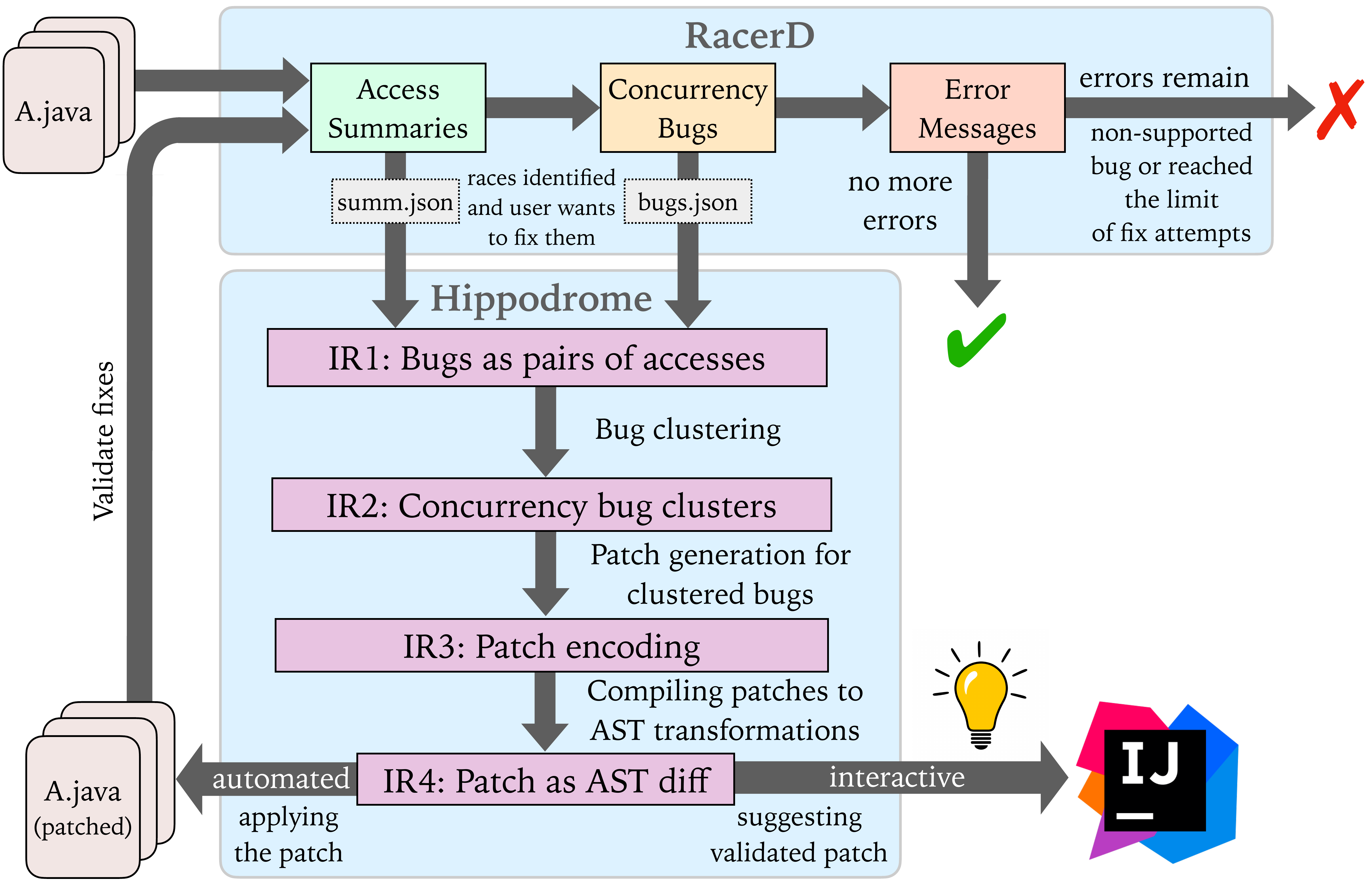}
	\caption{A summary of \tool workflow.}
	\label{fig:tool}
\end{figure}

\autoref{fig:tool} offers a bird's-eye view of our approach to data
race patch generation for Java programs.
An input program is first statically analysed for data races (\autoref{fig:tool}, top).
The access summaries and bugs collected from the analysis are merged into a
set of bugs, which are then clustered to support fixes across multiple related bugs.
We design a DSL to encode the patches before actually generating and
applying them to the input files.
The patches are repeatedly validated by the analysis (the back-link in
\autoref{fig:tool}) for the absence of deadlocks until no more alarms are
raised, in which case the bug is considered fixed and the patch is applied, if
an automated mode is chosen (bottom left).
In an interactive mode (bottom right), the fully validated patches are
suggested to the user via an IDE.

 We detail our choice of static analysis for fault localisation and the algorithmic 
 approach to patch synthesis in the following sections, highlighting what are the 
 abstractions required to connect these two components.


\section{Static Analysis Preliminaries}
\label{sec:analysis}

The analysis for fault localisation lies on top of  \racerd ~\cite{BlackshearGOS18},  
an open-source data race detector developed at Facebook. 
\racerd abides to the principles described earlier:
it is compositional and
it has been empirically demonstrated to provide high signal.    
We defer the tool's soundness and limitations discussion  to \autoref{sec:disc-limit}.
It now suffices to mention that the synchronisation primitives are limited to scoped locks and Java's \code{synchronized}
blocks (\ie, so-called \emph{coarse-grained} concurrency) ignoring,
\eg, atomic Compare-And-Set and any \emph{fine-grained}
synchronisation. 
This limitation turned out to 
be an advantage, since 
a specialised analysis reduces the number of false positives while performing well for industrial code where coarse-grained concurrency is the norm ~\cite{BlackshearGOS18}. 

According to textbook definitions, a race is caused by 
concurrent operations on a shared memory location, of which at least
one is a write~\cite{Herlihy-Shavit:08}.
Data races in object-based languages, such as Java, can be described
conveniently in a static sense (\ie, without referring to runtime
program state), in terms of the program's syntax, rather than memory,
via \emph{access paths}~\cite{jones1979flow}, which serve as
program-level ``representatives'' of dynamic operations with memory:
\begin{definition}[Syntactic data race]
\label{def:race}
Program statements $\stmt_1$ and $\stmt_2$ form a \emph{data race} if 
 two \emph{access paths}, $\spath_1$ and $\spath_2$ (\ie,
field-dereferencing chains $x.f_1.\ldots,f_n$, where $x$ is a variable or
\code{this}) pointing to the same memory location, 
are identified in $\stmt_1$ and $\stmt_2$, respectively, such that:
(a) executing $\stmt_1$ and $\stmt_2$ concurrently, they perform
two concurrent operations on $\spath_1$ and $\spath_2$, at least one
of which is a write, and (b) the sets of \emph{locks} held by
$\stmt_1$ and $\stmt_2$ at that execution point are disjoint.
  
\end{definition}

The analysis 
first conservatively detects Java classes
that can be used in a concurrent context, by checking for \code{@ThreadSafe}
annotations, locks, \code{synchronized} keyword, or if the instance of those
classes appear in the scope of a thread's \code{run()} method.\footnote{If a
  class intended to be used concurrently does not feature any of those
  ``indicators'', a developer might have to annotate it explicitly as
  \texttt{@ThreadSafe} to help the analysis.}
Subsequent stages work on the premise that \emph{any
  method} of such a class instance may have data races with \emph{any
  other method}, including itself.
The main routine operates on individual methods, deriving their
\emph{summaries}---exemplified below.
The analysis is compositional, as those summaries are later used for
analysing the code of the caller methods.
Finally, the analysis examines method summaries pair-wise in
an attempt to identify data races according to Def.~\ref{def:race}.

%


To identify potential races on access paths within concurrently
invoked methods of (the same instance of) a class $C$, the analysis
infers those methods' summaries comprised of \emph{access snapshots}.
An access snapshot contains mutual exclusion information used for
identifying possible data races.
\autoref{fig:abstract-domain-race} shows the analysis's abstract domain.
An access snapshot $\snapshot$ is a tuple
$\csummary{\spath}{k}{L}{L}{t}{o}{n}{\trace}$, where $\spath$ is the
access path, $k$ is a read/write indicator, $L$ abstracts the locks
protecting this access, $t$ and $o$ denote the thread kind and
ownership, respectively, and $\trace$ is the trace to $\spath$ and is
non-empty for indirect access (\ie, via method calls).
For brevity, we refer to elements of this tuple as, \eg,
$\snapshot.\spath$ for the access path in snapshot $\snapshot$.
The thread kind $t$ may take any of the three abstract values which form a partial order: 
$\mathrm{NoThread}$ - to denote that the current access snapshot belongs to a procedure which cannot run concurrently with any other threads, 
$\mathrm{AnyThreadButMain}$ -  when the access snapshot - belonging to the main thread - may be executed in parallel with background threads, 
or $\mathrm{AnyThread}$ - when the current access snapshot belongs to a procedure running on a thread that can interleave with any other thread.

A set $A_m$ of all snapshots collected for a method $m$ is, therefore,
an \emph{over-approximation} of all runtime heap accesses reachable
from $m$ via the corresponding syntactic access paths,
as well as of the corresponding locking patterns.

For the code in \autoref{fig:datarace}, the snapshots collected
by the analysis for the two \code{CustomerInfo} methods are as
follows (only showing the problematic access snapshots):

\noindent
{\footnotesize{
\[
  \begin{array}{@{\!\!}l}
   A_{\mathtt{withdraw}} = \left\{
    \begin{array}{l}
   \snapshot_{ \ref{line:1}}:~   
  \langle
  \mmcode{this.accounts[].balance}, {\rrd}, {0}, {\mathrm{AnyThread}}, 
 {{\unowned}, \{\mmcode{Account.setBalance()} \} 
  \rangle, }
  \\
   \snapshot_{ \ref{line:2}}:~
   \langle
   \mmcode{this.accounts[].balance}, {\rwr}, {0}, {\mathrm{AnyThread}}, 
   { {\unowned}, \{\mmcode{Account.getBalance()}\}  
   \rangle,} \ldots 
    \end{array}
  \right\}  
    \\[5pt]
    ~\\
   A_{\mathtt{deposit}} = \left\{
  \begin{array}{l}
  \snapshot_{ \ref{line:3}}:~   
  \langle
   \mmcode{this.accounts[].balance}, {\rrd}, {0}, {\mathrm{AnyThread}}, 
  {{\unowned},  \{\mmcode {Account.getBalance()}\}  
   \rangle, }
  \\  
  \snapshot_{ \ref{line:4}}:~
   \langle 
   \mmcode{this.accounts[].balance}, {\rwr}, {0}, {\mathrm{AnyThread}}, 
{ {\unowned}, \{\mmcode{Account.setBalance()}\} 
   \rangle,} \ldots      
    \end{array}
    \right\}
       \\[5pt]
    ~\\
    A_{\mathtt{getBalance}} = \left\{
    \begin{array}{l}
    	\snapshot_{ \ref{line:a1}}:~   
    	\langle
    	\mmcode{this.balance}, {\rrd}, {0}, {\mathrm{AnyThread}}, 
    {{\unowned},  \{\}  
    		\rangle }     
    \end{array}
    \right\}
       \\[5pt]
    ~\\
    A_{\mathtt{setBalance}} = \left\{
    \begin{array}{l}
    	\snapshot_{ \ref{line:a2}}:~   
    	\langle
    	\mmcode{this.balance}, {\rrd}, {0}, {\mathrm{AnyThread}},{{\unowned},  \{ \}  
    		\rangle }    
    \end{array}
    \right\}
  \end{array}
\]
}}

\begin{figure}[t]
\centering
\[
\!\!\!\!
{\small
\begin{array}{l@{\ }r@{\ }c@{\ }l@{\ }c@{\ \ }l}
  \text{access paths} & \spath & \in & \mathrm{Path} & = &
                                                           (\mathrm{Var}
                                                           \cup \set{\mcode{this}}) \times
                                                        \mathrm{Field^*} \\
   \text{trace}       & \trace & \in & T & = & (\mathrm{Class}
                                                     \times \mathrm{Meth})^*\\
  \text{locks} & L &\in & \mathcal{L} & = & \mathbb{N}\\
  \text{concurrent thds} & t & \in &\mathcal{T}  & =  &
                                                           \mathrm{NoThread} \sqsubset
                                                           \mathrm{AnyThreadButMain}
                                                     \\
  &&&&&
                                                           \sqsubset \mathrm{AnyThread}
  \\[2pt]                                     
   \text{ownership value} & o & \in &\mathcal{O} & = & \mathrm{\ownedif{\mathbb{N}}} \sqsubset
                                                    \mathrm{\unowned} 
  \\[2pt]
  \text{access type} & k & \in & \multicolumn{3}{@{}l@{\!\!}}{\set{\rrd,\rwr}}
  \\[2pt]
  \text{snapshot} & \snapshot& \in &\mathcal{A} & =
                                                   &\csummary{\spath}{k}{L}{L_s}{t}{o}{n}{\trace}
                                                     
  \\[2pt]
  \text{summary} & A &\subseteq& \mathcal{A} & = & \{ \snapshot \mid \snapshot \in \mathcal{A} \}
\end{array}
}
\]
\caption{Abstract domain for race detection.}
\label{fig:abstract-domain-race}
\end{figure}


The snapshot for line~\ref{line:1} states that there is a
\emph{read} access to a heap location pointed to by
\code{this.accounts[].balance}.
Moreover this memory location may be accessed by any
other thread ($\mathrm{AnyThread}$) unrestrictedly since there is no lock protecting
it (denoted via 0), and it is not owned by, i.e. not local to, any thread
($\unowned$).
The final element in the tuple represents the access's \emph{trace}
call, meaning that it is an indirect access done by calling the method
\code{setBalance()} of the class \code{Account}.
 
 Conflicts are detected in our running examples by pairwise checking
 against each other the snapshots in $A_{\mathtt{withdraw}}$ and
 $A_{\mathtt{deposit}}$, respectively.
 It concludes that every pair of paths refers to the same
 heap address and they may be accessed concurrently.
 Moreover the snapshot pairs
 $( \snapshot_{\ref{line:1}}, \snapshot_{\ref{line:2}} )$,
 $( \snapshot_{\ref{line:3}}, \snapshot_{\ref{line:4}} )$,
 $( \snapshot_{\ref{line:1}}, \snapshot_{\ref{line:4}} )$ and
 $( \snapshot_{\ref{line:2}}, \snapshot_{\ref{line:3}} )$ involve at
 least one write. 

\section{Inferring and Applying Patches}
\label{sec:repair}

This section describes the enhancements we brought to the static analysis, 
what is the space of solution, and how we finally infer and apply the patches.

\subsection{Fault Localisation}
\label{sec:localization}

We first motivate and describe the enhancements of the static analysis for fault localisation.

\subsubsection{Bug reporting strategy} \label{sec:bug-report}

To reduce the number of warnings which may be overwhelming for the developer,
\racerd only reports at most one pair of read-write conflicting snapshots per
access location.
For instance, it will report
$(\snapshot_{\ref{line:1}}, \snapshot_{\ref{line:2}})$,
$ ( \snapshot_{\ref{line:3}}, \snapshot_{\ref{line:4}} )$ as
potential bugs for the example  in  \autoref{fig:datarace}, although
as mentioned in the previous section there are two more problematic pairs,
namely
$( \snapshot_{\ref{line:1}}, \snapshot_{\ref{line:4}} )$ and
$( \snapshot_{\ref{line:2}}, \snapshot_{\ref{line:3}} )$.
This is a sensible design choice for manual repair where the
developer is directed progressively to new conflicts as she repairs
old ones and re-runs the analysis.
Automatic repair on the other hand can be instrumented to process
multiple conflicts at once.
A complete view of the bugs allows for a better patch generation strategy, hence
we modify the analysis to log all possible bugs.

\subsubsection{Enhanced lock domain}
\label{sec:domain}

A na\"{i}ve (and wrong) fix for the race in \autoref{fig:datarace}
is to wrap lines \ref{line:1}-\ref{line:2} in
\code{synchronized(m1)\{} \ldots \code{ \}}, and lines \ref{line:3}-\ref{line:4}
in \code{synchronized(m2)\{}\ldots\code{\}}, where \code{m1} and \code{m2} are freshly created mutexes such that 
 \code{m1} $\neq$ \code{m2}.
The access snapshots would be the same except for the locks
component, which would change from 0 to 1 in each of the four tuples, making
\racerd erroneously infer the \emph{absence} of a race, due to the
unsoundness caused by non-conservative interpretation of its abstract
domain $\mathcal{L}$ for locking
(\cf~\autoref{fig:abstract-domain-race}).
This is because its lock domain is defined to only track the
\emph{number} of locks used to protect an access in the actual code,
but not their identity.
As a remedy to this unsoundness, we change the abstract domain of
locks $\mathcal{L}$ from natural numbers to the \emph{powerset of
	paths} (\ie,~the new $\mathcal{L} \eqdef \wp(\mathrm{Path})$), where the
identity of each involved lock is abstracted by its syntactic access
path.
The new domain affords better race detection, while offering crucial
information for the repair purposes: {\emph{which} exact locks are
	taken at the location of a race}.
In the remainder of the paper, we will keep referring to the enhanced
analysis implementation as \racerd.

\subsubsection{Static Race Detection from Summaries}

 A data race is defined as a relation on two snapshots:
%


{
		\[
		\begin{array}{l@{\!\!}ll}
			\multicolumn{2}{l}{\letz~ \race(a_1,~ a_2) ~=~ }&
			\\
			~~ & ~~~ a_1.\spath = a_2.\spath & \textit{(same path)}
			\\
			& \wedge~ (a_1.k = \rwr \vee a_2.k = \rwr) & \textit{(at least one write)}
			\\
			& \wedge~ a_1.L  \cap a_2.L  = \emptyset  & \textit{(disjoint locks)}
			\\
			& \wedge~ a_1.t \sqcup a_2.t = \mathrm{AnyThread} & \textit{(possibly concurrent)}
			\\
			& \wedge~ (a_1.o  \, = \,  \mathrm{Unowned} \wedge a_2.o \, = \, \mathrm{Unowned}) & \textit{(shared
				memory)}
		\end{array}
		\]
}
%


\noindent
That is, two methods, $mth_1$ and $mth_2$ manifest a race if there exist snapshots
$a_1\in A_{mth_1}$ and $a_2 \in A_{mth_2}$ such that $\race(a_1,a_2) = \True$.
Two accesses are, thus guaranteed to not form a data race if they are both
protected by the same lock or performed by the same thread.

%
In addition to the binary data race formulation, we define a unary
relation over access snapshots as follows:
%
%
{
		\[
		\begin{array}{lll}
			\multicolumn{2}{l}{\letz~ \wrace~ a ~=~ }&
			\\ ~~ &  a.k = \rwr  & \textit{(write operation)}
			\\ &\wedge~ a.L = \emptyset  & \textit{(unlocked)}
			\\ & \wedge~ a.t = \mathrm{AnyThread}  & \textit{(possibly concurrent)}
			\\ & \wedge~ a.o = \unowned  & \textit{(unowned resource)}
		\end{array}
		\]
}

\noindent
A method $mth$ is said to be \emph{unsafe} when there exists
$a \in A_{mth}$ such that $\wrace(a) = \True$.
In the next section, we show how the enhanced domain along with the
definition of a data race and unsafe method suffice to derive a patch,
and subsequently a fix for the example from \autoref{fig:datarace}.
\subsection{Searching for Repairs}
\label{sec:search}

Let us revisit the running example in \autoref{fig:datarace} with
synchronisation missing.
In particular, let us focus on the bug described by the pair 
$(\snapshot_{\ref{line:1}}, \snapshot_{\ref{line:2}})$. 
With no lock whatsoever to protect these accesses we are in the
situation where a number of fixes can be applied in a possibly
automated way:
\begin{itemize}
	\item We can protect both accesses by individually wrapping the
affected lines into \code{synchronized(this)} statements. While
straightforward, relying on these so-called intrinsic locks is against
best practices in concurrent programming due to introduced
synchronisation overheads.
\item Alternatively, we can create a fresh object to add the
two \code{synchronized} statements on it, thus avoiding contention with
other \code{synchronized} statements in this class. This approach may
generate too many locks, especially for accesses involved in
multiple data races.
\item Finally, we can annotate the shared variable as
\code{volatile}. 
While many developers' favourite solution, best practices only
recommend it when the resource is shared for reading by
multiple threads, but written to by only one ~\cite{jcip}---a
piece of knowledge beyond the reach of static analysis which detects pairs of program components that may be executed concurrently 
but not the number of threads to be spawned for a particular code.
 Furthermore,
this approach is not straightforward to apply for array entries.
\end{itemize}

This short informal analysis 
highlights the
challenges in choosing a strategy for automatic patch generation, with
the observation that it is often a compromise between simplicity of a
fix and avoiding over-synchronisation.
Since over-synchronisation is highly probable when
using intrinsic locks exclusively, we chose to avoid them in our
approach and focus mostly on a \emph{careful choice of mutexes for
  manipulating \code{synchronized} blocks}.

Although the above-mentioned solutions are common in APR or manual repair, 
none tackles the root cause of the
race, namely that the accesses to \code{this.accounts[].balance} are
unprotected in the \code{Account} class (which is possibly used
somewhere else, too).
This observation shows that treating a concurrency bug as a standalone
entity would deprive one from opportunities to produce high-quality
fixes for families of related issues.
To regain the holistic view on the origins and the implications of
data races, we propose a mechanism to \emph{cluster related bugs}
allowing us to find a suitable common patch.

\subsection{Patch Inference}
\label{sec:localization}

The main pipeline of \tool transforms summaries and bug reports
obtained from the bug analysis (\autoref{fig:tool}, top) to a sequence of
Intermediate Representations (IRs), eventually obtaining patch
candidates for the discovered concurrency bugs. 
We proceed to describe the involved IRs and the corresponding
algorithms for producing them.

{\bf IR1.} A bug is reported either as
a pair of access snapshots (in the case of \rwbug ~races) or as a
single snapshot (in the case of unprotected writes):
%

{
	\centering
	\[
	\!\!\!\!
	{
		\begin{array}{l@{\ \ }l@{\ \ }c@{\ \ }l}
			\text{bug} & \bug \in \bugsad & = &  
			\{ \langle \snapshot \rangle \mid \wrace(\snapshot)  = \True  \} ~\cup~
			\{ \langle \snapshot_1,  \snapshot_2\rangle \mid \race(\snapshot_1, \snapshot_2)  = \True  \}
		\end{array}
	}
	\]
}

%
\noindent

Given a \rwbug ~bug $b$, a straightforward strategy is to
pick any one of the locks in $b.a_1.L\,{\cup}\,b.a_2.L$ if any exists, or
create a fresh one otherwise.
For instance, for the bug
$( \snapshot_{\ref{line:1}}, \snapshot_{\ref{line:2}} )$ in our
example (\autoref{fig:datarace}), since
$\snapshot_{\ref{line:1}}.L\, {\cup}\, \snapshot_{\ref{line:2}}.L =
\emptyset$, we would create a new object, say \code{m1}, and insert
the appropriate \code{synchronized} blocks.
Similarly, we could fix the $( \snapshot_{\ref{line:3}}, \snapshot_{\ref{line:4}} )$ via a fresh mutex, say \code{m2}.  
For the remaining two bugs which involve inter-method accesses we would  create new mutexes, or choose from the freshly created ones. 
Either way this solution already creates nested synchronisation blocks breaking maintainability and increasing the chance of deadlocks. 

\noindent\emph{Example 1.} To better highlight these issues consider in \autoref{fig:naive-fix} a simplified, yet more general instance of the above situation depicting three program statements sharing the same resource ${\spath}$, where 
at  least two of the statement involve a write operation on ${\spath}$.
\noindent
Initially, \code{s}$ _2$ and \code{s}$_3$ are protected by mutexes \code{m1} and \code{m2}, respectively. 
The analysis would report   
$( \snapshot_1 ,\snapshot_2 )$,
$( \snapshot_2, \snapshot_3 )$ and
$( \snapshot_1, \snapshot_3 )$ as data races. 
Zooming individually into each bug could lead to fixes where \code{synchronized} statements are introduced as suggested by the arrows. 
The arrows are annotated with the order in which the fix is generated. 
Clearly this pseudo-fix is problematic, not only due to the clutter of  synchronisations but also because it  na\"{i}vely introduced a deadlock (dashed arrow). 
Can we do better?

\begin{figure}
{
    \newcounter{tmkcount}
	\tikzset{
		use tikzmark/.style={
			remember picture,
			overlay,
			execute at end picture={
				\stepcounter{tmkcount}
			},
		},
		tikzmark suffix={-\thetmkcount}
	}

\noindent
\hspace{0.5cm}
\begin{minipage}{.25\textwidth}
	\begin{lstlisting}[title=Access $a_1$,mathescape=true,name=code1]
		
	//synchronized(m1){
	//  synchronized(m2){
	    	 s$_3(\spath)$
   //} }   
	\end{lstlisting}
\end{minipage}%
 %
 %
 \hfill
\begin{minipage}{.25\textwidth}
	\begin{lstlisting}[title=Access $a_2$,mathescape=true,escapechar=\%,name=code2]
		
		
 synchronized(m1){
     s$_2(\spath)$
 }  
	\end{lstlisting}
\end{minipage}%
%
%
\hfill
\begin{minipage}{.25\textwidth}
	\begin{lstlisting}[title=Access $a_3$,mathescape=true,name=code3]
		
	synchronized(m2){
	//  synchronized(m1){
	       s$_3(\spath)$
   } //} 
	\end{lstlisting}

\end{minipage}%
	\begin{tikzpicture}[use tikzmark]
		\coordinate (aux1s)  at (pic cs:line-code2-3-end);
		\coordinate (aux1e) at (pic cs:line-code3-3-start);
		
		\coordinate (aux2s)  at (pic cs:line-code2-3-start);
		\coordinate (aux2e) at (pic cs:line-code1-2-end);
		
		\coordinate (aux3s)  at (pic cs:line-code3-2-start);
		\coordinate (aux3e) at (pic cs:line-code1-3-end);
	\end{tikzpicture}

    \begin{tikzpicture}[use tikzmark]
    	\draw[->] ([yshift=0.5ex,xshift=0.5ex]aux1s) to 
    	node[midway,above,sloped]{\tiny(2)}
    	([yshift=0.5ex,xshift=1ex]aux1e);
    	
   		\draw[->] ([yshift=0.7ex,xshift=0ex]aux2s) to[bend right=10]  	
   		node[midway,above,sloped]{\tiny(1)}
		([yshift=0.7ex,xshift=1ex]aux2e) 	;  		
   		
    	\draw[dashed,->] ([yshift=0.8ex,xshift=0ex]aux3s) 
    	to[bend right=7]  
    	node[midway,above,sloped]{\tiny(3)} 
    	([yshift=0.7ex,xshift=1ex]aux3e) 	;
    \end{tikzpicture}
}
\caption{Three data races and their na\"{i}ve, order-sensitive fixes.}
\label{fig:naive-fix}
\end{figure}

{\bf IR2.} 
To improve patch quality, instead of deriving patches from
individual bugs, we will infer them from bug \emph{clusters}. We will
do so by analysing \emph{sets} (instead of \emph{pairs}) of snapshots,
grouping bugs according to their shared access paths.
For a set of bugs $B$, a cluster $\mathcal{C}_B$ is defined as
follows:
\begin{center}
{
${\mathcal{C}_B {\eqdef}
	\set{
	\bugset' {\subseteq} \bugset \left| 
	\exists \pi',
	\forall b' { \in}  B', 
	 b  {\notin}  B',
	\!\!\!\!\! {\bigcup\limits_{\snapshot \in \acc(b')} }\!\!\! \snapshot . \spath  {=} \{  \pi' \}  
	\wedge
	 \pi'  {\notin} \!\!\! {\bigcup\limits_{\snapshot \in \acc(b)} }\!\!\! {\snapshot . \spath}
    \right.}
}   $ 
}
\end{center}

\vspace{2pt}

\noindent where the function $\acc$ returns the set of snapshots
summarising the bug.
For the clarity of presentation, we assume that all the access snapshots
in a cluster belong to the same class (in practice, we refine the
above clustering by taking classes into the account).
We define a function $\cls(a)$ to return the class of an access
snapshot $a$, and generalise it to indicate the class of a bug cluster
(since all its bugs share the same access).

For our running example, a cluster of bugs related by their unprotected access
to  \code{this.accounts[].balance} is the set \\
\centerline{$\{( \snapshot_{\ref{line:1}}, \snapshot_{\ref{line:2}} ),
( \snapshot_{\ref{line:3}}, \snapshot_{\ref{line:4}} ),
( \snapshot_{\ref{line:1}}, \snapshot_{\ref{line:4}} ) 
( \snapshot_{\ref{line:2}}, \snapshot_{\ref{line:3}} ) \} $}

\textbf{IR3.} 
Algorithm~\ref{alg:main} infers patches from bug clusters.
%
Patches are encoded using the following simple domain-specific
language which supports insertion of \code{synchronized} blocks, variable
declarations, and \code{volatile} annotations:
\[
{\small
	\begin{array}{l @{\ }r@{\ \ }c@{\ \ }l}
		\text{action} ~~~& \idn{act}  & ::= & {\textsc{SYNC}(A, \idn{lock}) \mid \textsc{DECLARE}(\idn{class},\idn{a},\idn{x},\idn{t}) } 
                        \mid ~ {\textsc{VOLATILE}(\snapshot) }   ~\mid~
                             \textsc{NIL} 
		\\ 
		\text{composition~~} & \patch & ::= & \idn{act} ~\mid~
                                          {\textsc{AND}(\idn{act}, \idn{act}) ~ \mid ~ \textsc{OR}(\idn{act},\idn{act})}
	\end{array}
}
\]


%
\textsc{AND} is used to compose patch components corresponding to
different snapshot accesses of the same bug cluster. 
\textsc{OR} denotes different patch options for the same access
snapshot.

\begin{algorithm}[t]
	\caption{\algo{CreatePatchEncodings}}
	\label{alg:main}
	\textbf{Input}:  ~~a powerset of bugs  $\idn{\mathcal{C}}$\\
	\textbf{Output}: a patch set $\idn{\patchset}$\\
	$P_{\idn{\mathcal{C} }}\leftarrow \emptyset$\\ 
	\For {${\idn{\bugset} \in \mathcal{C}}$} {
		$A \leftarrow$ union of $\idn{acc(b)}$ for all $b$ in $\bugset$
    \acomment{ // snapshots for a cluster $B$}\\
		${\idn{locks} \leftarrow}$ union of $\snapshot.L$  for all $a$ in $A$\\
		\uIf {$\idn{locks} = \emptyset$} { 			{\acomment{ // if no locks available, create a new mutex }}\\
			{$\idn{var} \, \leftarrow $ fresh variable name} \\
			{$\idn{act} ~\, \leftarrow {\textsc{DECLARE}(\idn{cls}(B),\idn{var},\idn{Object})}$}\\
			$\idn{locks} \leftarrow \{\idn{var}\}$\\    
		}
		\uElse {
			$\idn{locks} \leftarrow $ order $\idn{locks}$ according to their frequency (descending)
			 \\
			$\idn{act} \leftarrow \textsc{NIL}$\\
		}
       $\idn{fixes} \leftarrow \textsc{NIL} $ {\acomment{// collects all possible  fixes for cluster $\bugset$}}  \\
	   \For {${\idn{lock} \in \idn{locks}}$} {
			${\patchset \leftarrow \emptyset}$ {\acomment{// collects patch components}} \\
			\For {${\snapshot \in A}$} {
				$a' \leftarrow$ find the innermost access snapshot in
        $a.\trace$ common to all accesses in A
        \\
				{\acomment{ // synchronising the statement containing $a'$ via
          $\idn{lock}$}}
        \\
				$p \,\,\leftarrow \textsc{SYNC}(\{a'\}, \idn{lock})$\\
				$P \leftarrow \{p\} \cup P$ }
			$\idn{patch} \leftarrow \textsc{AND}(\idn{act},
      \textsc{AND}(P))$ 			{\acomment{// combine patch components}}\\
			$\idn{fixes} \leftarrow \textsc{OR}(\idn{fixes},
			\idn{patch})$ 
       }
		$ x ~~ \leftarrow$ a field subject of race in all bugs of $B$\\
		$\patch_B \leftarrow \textsc{OR}( \idn{fixes},$ \textsc{VOLATILE}$(x, \idn{cls}(B)) )$  \\
		$P_{\mathcal{C}}  \leftarrow \{\idn{\patch_B}\} \cup P_{\mathcal{C}}$	\\

		}
	return $P_{\mathcal{C}}$ 
\end{algorithm}
\setlength{\textfloatsep}{4pt}

Even though seemingly simple, Algorithm~\ref{alg:main} has a few 
subtle points.
First, for each bug cluster, the algorithm derives a patch to ensure
that all of its accesses are protected by a common lock.
As far as we know, there is no best strategy for choosing this lock
when some access paths are already protected.
Of several possible strategies applicable in this case, we have
settled on choosing a lock that protects the most accesses, thus
avoiding nested synchronisation.
However, the lock choice strategy can easily be adjusted by choosing 
how to manipulate the list of locks at line 13. 
Since a final patch is a composition of sub-patches (line 23), the algorithm could easily 
support a more advanced  
lock choice strategy (if one would be proposed in the future) 
which could introduce multiple mutexes for a cluster of bugs as opposed to a common mutex for all bugs in a cluster as currently proposed.    

Another subtle point of the algorithm is protecting the innermost
access in the call chain leading to a race, thus solving the root
cause of the bug and reducing the number of locations to be
synchronised.
An alternative solution is to synchronise at the call site.
We support this option too, though not as a default.
The benefit of this alternative is that it could be tailored to solve
simple atomicity violations~\cite{LuciaDSC08}, but those are not our
main focus.

The recommended patch for the running example is as follows (written in infix order for readability, and ignoring the volatile solution for brevity):

\centerline{$
	\textsc{DECLARE}(\idn{Account},\idn{v},\idn{Object}) ~
		\textsc{AND}~
	\textsc{SYNC}(\{\snapshot_{getBalance}\},\idn{v})
	~\textsc{AND}~
	\textsc{SYNC}(\{\snapshot_{setBalance}\},\idn{v})
  $}

\noindent where $\snapshot_{getBalance}$ and $\snapshot_{setBalance}$ are the  snapshots corresponding to the accesses to \code{this.accounts[].balance} from \code{Account}'s  \code{getBalance} and  \code{setBalance}, respectively. In other words, \tool recommends patching a bug at its source instead of the location of its manifestation.

{\bf IR4.} 
The final step in the \tool pipeline is to translate  the
encodings from the preceding stage to actual patches.
This process is described by Algorithm~\ref{alg:patch}, which produces
patches in the following simple language of AST-manipulating actions
taking tree nodes as their arguments:


\[
\!\!\!\!
{\small
	\begin{array}{@{\!\!\!} l @{\!\!}r@{\ \ }c@{\ \ }l}
		\text{action} ~~~& \fact  & ::= & ~~{\textsc{REPLACE}(\idn{from}, \idn{to}, \idn {ast}) }\\[2pt]
	    	&&& \mid \textsc{INSERT\_AFTER}(\idn{stmt},\idn{ins},\idn{ast}) \\[2pt]
			&&& \mid \textsc{INSERT\_BEFORE}(\idn{stmt},\idn{ins},\idn{ast}) \\[3pt]
		\text{composition~~} & \fpatch & ::= & \fact ~\mid~ {\textsc{AND}(\fact,\fact) ~ \mid ~ \textsc{OR}(\fact,\fact)}
	\end{array}
}
\]


\begin{algorithm}[t]
	\caption{\algo{CreatePatch}}
	\label{alg:patch}
	\textbf{Input}:  ~~a patch encoding $\idn{\patch}_0$ in IR3  format\\
	\textbf{Output}: a patch as AST diff ${\idn{\fpatch}}$\\
	$\idn{\patch}  \leftarrow $ normalise \textsc{AND} such that any $\textsc{SYNC}(\idn{A_1,lock})$ and $\textsc{SYNC}(\idn{A_2,lock})$ are merged into  
	$\textsc{SYNC}(\idn{A_1 \cup A_2,lock})$ if  all the snapshots in 
	$A_1$ and $A_2$ belong to the same method\\
	\Switch{$\idn{\patch}$}{
		\Case{$\textsc{AND}(\idn{\act_1, \act_2})$}{
		return $\textsc{AND}(\algo{CreatePatch}(\idn{\act_1}),\algo{CreatePatch}(\idn{\act_2}))$}
		\Case{$\textsc{OR}(\idn{\act_1, \act_2})$}{
		return $\textsc{OR}(\algo{CreatePatch}(\idn{\act_1}),\algo{CreatePatch}(\idn{\act_2}))$}
		\Case{$\textsc{SYNC}(\idn{A,lock})$}{
		return $\algo{InsertLock}(\idn{\patch})$}
		\Case{$\textsc{DECLARE}(\idn{class,x,t})$}{
		return $\algo{DeclareVariable}(\idn{class,x,t})$}
		\Case{$\textsc{VOLATILE}(\idn{a,x,t})$}{
		return $\algo{MakeVolatile}(\idn{\patch})$}
	}
\end{algorithm}
 
The algorithm relies on several auxiliary procedures for declaring
variables, adding volatile modifiers, and inserting locks.
We omit the definitions of all but the
last one, as the rest are straightforward.
We note that the only delicate matter for \algo{DeclareVariable} is to
identify whether any of the \textsc{SYNC} accompanying the peer
\textsc{DECLARE}s belongs to a static method.
If that is the case the declared variable must be annotated
as \code{static}, too.
Algorithm~\ref{alg:insertLock} defines the insertions of
\code{synchronized} blocks, which might require splitting variable
definitions into declarations and initialisations.

In our running example, the algorithm inserts a variable declaration, instantiates an object in the \code{Account} class,
and replaces in 
\code{getBalance} and  \code{setBalance} the two unprotected accesses to \code{this.accounts[].balance}   with their corresponding accesses protected by the freshly created object. 
For \emph{Example~1}, the fix produced by \tool avoids the deadlock by only inserting 
synchronisations (1) and (2).

\begin{algorithm}[t]
	\caption{\algo{InsertLock} \label{alg:insertLock}}
	\textbf{Input}:  ~~a \textsc{SYNC}$({A}, \idn{lock})$ patch\\
	\textbf{Output}: a patch as AST diff ${\idn{\fpatch}}$\\
	$\idn{class}~\leftarrow $ retrieve AST of $\idn{\cls(A)}$\\
	$\idn{from}~\leftarrow $ the closest parent in $\idn{class}$ of a node containing all $\idn{\snapshot}$ in $A$ \\
	$\idn{cond}~\leftarrow$ true if $\idn{from}$ has variables that
  outlive the scope of $\idn{from}$ \\
	\If{$\idn{cond}$}{
		$\idn{decl}, \idn{from}' \leftarrow $ move variable declarations
    out of $\idn{from}$ \\
    \acomment{// $\idn{from}'$ now contains only initialisers of the variables}
    \\
		$\fact_d \,\leftarrow $  \textsc{INSERT\_BEFORE}($\idn{from,decl,class}$)\\
		$\idn{sync} \leftarrow $ wrap a \lstinline[language=Java]{synchronized}~(\emph{lock}) \lstinline{\{\}} around $\idn{from}'$\\
		$\fact_i  ~ \leftarrow$  \textsc{REPLACE}($\idn{from,sync,class}$) \\
		return \textsc{AND}$(\idn{\fact_d,\fact_i})$		
	}\Else{
		$\idn{sync} \leftarrow $ wrap a
    \lstinline[language=Java]{synchronized}~(\emph{lock}) \lstinline{\{\}} around $\idn{from}$\\
	   $\idn{\fact ~\,\, \leftarrow}$  \textsc{REPLACE}($\idn{from,sync,class}$) \\
	   return $\idn{\fact}$
   }
\end{algorithm}

The resulting patches are normalised into a ``disjunctive normal form''.
When \tool is used in an interactive mode, the user receives a list of
complete patches to choose from.
When used in auto-repair mode, the tool favours the \textsc{SYNC}
patches with the least number of insertions.

\paragraph*{Patch application and validation}
\label{sec:applying-patches}

The patch application is straightforward, involving changes in the
original file's AST.
Given a fix $\textsc{OR}(\fact_1,\fact_2)$  for a bug cluster $\bugset$, the patch application strategy chooses to apply patch $\fact_1$, and stores $\fact_2$ for subsequent application should the validation phase fail. 
Once completed, the files are validated by \racerd (back-link in
\autoref{fig:tool}).
The analysis \emph{can detect deadlocks}---a feature we actively use to
discard patches that might introduce such issues.
Even with no deadlock detected, the program may still contain bugs (\eg,
data races on a different access path), hence we keep the fix and
reiterate through the algorithm to fix the remaining issues, doing so
until there are no more bugs left, or the iteration limit (currently 10) is
reached.
In our case studies, we have never hit that limit.
We guarantee to introduce \emph{no new races} due to the fact that we
never update existing locks, but only introduce more synchronisation.
In \autoref{sec:disc-limit}, we state our formal correctness claims, enumerating
the assumptions under which they hold.

%

\paragraph*{Patch quality}

In designing the strategy for generating and automatically choosing
patches, we have considered reducing both (a) context switching as well as (b)
 over-synchronisation.
We optimise for (a) by clustering bugs based on their shared access
path and by merging lock statements on the same mutex located in the
same method (the latter one is optional).
We optimise for (b) by avoiding intrinsic locks (\ie, on \code{this}), and by
optimising each bug cluster to be served by mutexes not serving to synchronise
any other clusters.


\vspace{-5pt}

\section{Implementation and Evaluation}

\subsection{Implementation}
\label{sec:impl}

\setlength{\textfloatsep}{15pt}

As per  \autoref{fig:tool}, our approach is implemented in a framework comprising the following components: \emph{bug detector}, \emph{patch synthesiser} and \emph{application}, and \emph{validator}.
For \emph{bug detection} and \emph{validation}\footnote{While working on the validation phase, we discovered a bug in how the locks were treated in the deadlock detector which we have fixed in our analysis and responsibly reported it to the \inferr team.}, we build on {\racerd} (implemented in OCaml)  
extending its lock domain (\autoref{sec:domain}) and the bug reporting strategy (\autoref{sec:bug-report}).   
\racerd already stores concurrency bug reports in log files, but not
the method summaries, for which we had to add support.
The interaction with \tool's core components is achieved via
JSON files.
\tool's \emph{patch synthesiser and application} are implemented in Scala, comprising of about 2.5k lines of code.
The synthesiser expects as input a set of bugs and a set of method summaries, and then proceeds to construct all the intermediate representations (\autoref{sec:localization}).   
Although we have only considered patches which involve inserting  \code{synchronized} statements or  \code{volatile} annotations, staging the patch synthesis process into IRs allows future works to add support for other synchronization primitives by simply extending these IRs correspondingly. 
Manipulation with the Java files when constructing patches
(\autoref{sec:repair}) is done via ANTLR v4.

When \tool runs in automatic mode, multiple patches may be generated for a single cluster of bugs. 
To decide which patch is the best fit, we have added parametrised support for a cost function to choose the best fix candidate.
We have implemented a simple cost function which measures the number of statements which need to be updated for a fix, and choose the least intrusive fix (smallest number of updates). 
Users of our framework could explore other application specific cost functions if the default cost function is not satisfactory. 
Only the best candidate fix is validated in this mode.

\begin{figure}[t]
\setlength{\belowcaptionskip}{-10pt}
 \includegraphics[width=\textwidth]{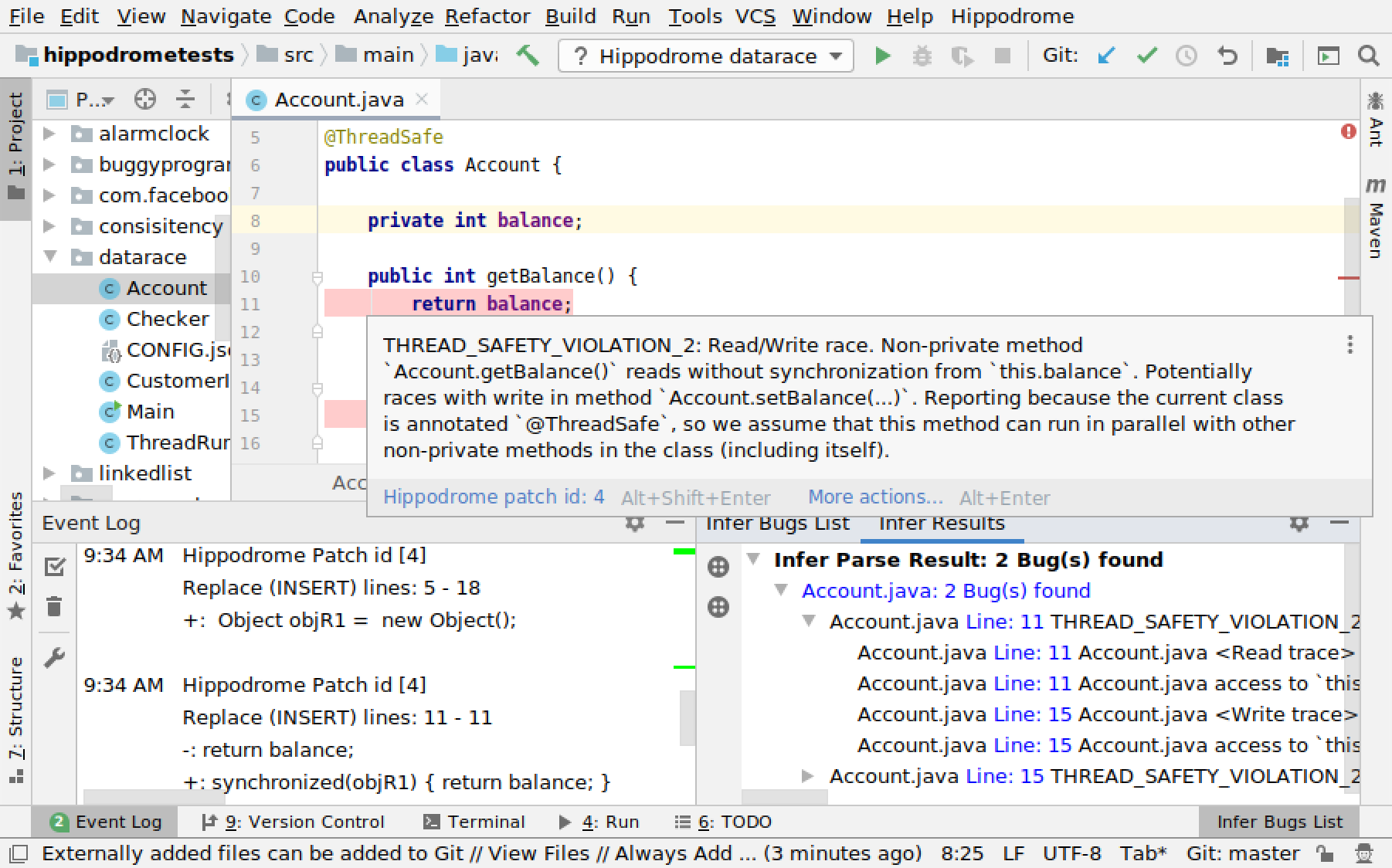}
\caption{Using \tool in an interactive mode.}
\label{fig:intellij}
\end{figure}

When \tool runs in interactive mode, all possible fixes are validated before being presented to the user which then chooses the fix it considers as best fit. 
For a better UX experience, we have integrated \tool into IntelliJ IDEA as a plug-in.
The screenshot in \autoref{fig:intellij} demonstrates a report
produced for a data race in the 
\code{Account} class, as well as a patch suggested by \tool (bottom left
window), which will create a new mutex object \code{objR1} and use it
for synchronising accesses to \code{Account}'s field \code{balance}.

Since we have kept a clean separation between the bug analysis and the implementation of the patch generation mechanism, the framework could easily benefit from the advancements in data race detection, e.g. \cite{Liu2021}, as long as the artefacts of these new technologies can discharge to JSON files the analysis results in the format described in \autoref{fig:abstract-domain-race}. 

Our tool \tool is publicly available as a Docker image~\cite{racerdfix}, or as an open-source project~\cite{hippodrome}.

\begin{table*}[]
\caption{Results on subject apps by \pfix and  \tool. All times are given in seconds. Det. means ``detection''.}
\label{tab:tab_rq1}
\begin{adjustbox}{width=\linewidth}
\centering
\begin{tabular}{lcl|ccc|c@{\ }|ccc}
\toprule
\multicolumn{3}{c|}{\textbf{Subject Apps}}                              & \multicolumn{3}{c|}{{\pfix}}                                & \multirow{2}{*}{\!\!\racerd \textbf{Det.}} & \multicolumn{3}{c}{{\tool}}                           \\[2pt] 
\textbf{Program Name} & \textbf{LOC} & \textbf{From Benchmark Suites} & \textbf{Fix Status} & \textbf{Det. Time} & \textbf{Fix Time} &                                            & \textbf{Fix Status} & \textbf{Det. Time} & \textbf{Fix Time} \\ \midrule
account               & 102             & PECAN, JaConTeBe, PFix         & Success             & 22.7                    & 238.75            & Fail                                       & N/A                  & N/A                      & N/A                \\ 
accountsubtype        & 138             & PFix                           & Success             & 29.4                    & 21.4              & Fail                                       & N/A                  & N/A                      & N/A                \\ 
airline               & 51              & JaConTeBe, PFix                & Success             & 8.35                    & 16.2              & Success                                    & Success             & 1.34                    & 0.71              \\ 
alarmclock            & 206             & JaConTeBe, PFix                & Fail                & 10.75                   & N/A           & Success                                    & Success             & 4.51                    & 2.07              \\ 
allocation vector     & 114             & JaConTeBe                      & Fail                & N/A                      & N/A                & Success                                    & Success             & 2.42                    & 0.73              \\ 
atmoerror             & 48              & PFix                           & Success             & 7.3                     & 5.95              & Success                                    & Success             & 1.45                    & 0.43              \\ 
buggyprogram          & 258             & PECAN, PFix                    & Success             & 9.45                    & 33.55             & Partial                                    & Partial             & 1.61                    & 2.87              \\ 
checkfield            & 41              & PFix                           & Success             & 7.15                    & 9.8               & Success                                    & Success             & 1.24                    & 0.8               \\ 
consisitency          & 28              & PFix                           & Success             & 6.75                    & 9.95              & Success                                    & Success             & 2.3                     & 0.87              \\ 
critical              & 56              & PECAN, JaConTeBe, PFix         & Success             & 15.4                    & 14.1              & Success                                    & Success             & 5.8                     & 1.15              \\ 
datarace              & 90              & PFix                           & Success             & 8.1                     & 51.15             & Success                                    & Success             & 1.28                    & 0.28              \\ 
even                  & 49              & PFix                           & Success             & 7.25                    & 91.15             & Success                                    & Success             & 2.14                    & 0.36              \\ 
hashcodetest          & 1,258            & PFix                           & Success             & 8.45                    & 7.45              & Success                                    & Success             & 4.62                    & 1.65              \\ 
linkedlist            & 204             & PECAN, JaConTeBe, PFix         & Success             & 7.95                    & 35.25             & Success                                    & Success             & 0.97                    & 1.11              \\ 
log4j                 & 18,799           & JaConTeBe, PFix            & Success                 & 22.9                    & 20.35             & Success                                     &
    Success           & 1.49                      & 1.68               \\
Manager               & 130             & PECAN                          & Fail                & N/A                      & N/A                & Success                                    & Success             & 2.54                    & 1.43              \\ 
mergesort             & 270             & PECAN, PFix                    & Fail                & 17.95                   & N/A             & Success                                    & Success             & 1.07                    & 2.87              \\ 
pingpong              & 130             & PFix                           & Success             & 25.2                    & 23.05             & Success                                    & Success             & 3.67                    & 1.48              \\ 
ProducerConsumer      & 144             & PFix                           & Fail                & 16.0                      & N/A             & Success                                    & Success             & 4.61                    & 1.61              \\ 
reorder2              & 135             & JaConTeBe, PFix                & Success             & 7.7                     & 11.9              & Success                                    & Success             & 1.32                    & 0.58              \\ 
store                 & 44              & PFix                           & Success             & 7.2                     & 5.85              & Success                                    & Success             & 1.22                    & 0.29              \\ 
stringbuffer          & 416             & PECAN, PFix                    & Success             & 7.0                       & 22.2              & Fail                                       & N/A                  & N/A                      & N/A                \\ 
wrongLock             & 73              & JaConTeBe, PFix                & Success             & 7.15                    & 5.9               & Success                                    & Success             & 1.24                    & 0.36              \\ 
wrongLock2            & 36              & PFix                           & Success             & 7.3                     & 16.4              & Success                                    & Success             & 1.39                    & 0.94              \\ 
\bottomrule
\end{tabular}
\end{adjustbox}
\end{table*}


\subsection{Evaluation}
\label{sec:eval}
\setlength{\textfloatsep}{15pt}

We empirically evaluated \tool's effectiveness in producing high-quality
fixes for Java data races. Experiments were designed to answer the
following Research Questions:

\begin{description}

\item \textbf{RQ1:} How does \tool compare to the state-of-the-art
  repair tools in terms of performance and efficacy?

\item \textbf{RQ2}: What is \tool's performance on large projects and
  how do the patches it produces compare to developers' manual fixes in
  those projects in terms of quality?

\end{description}

\noindent
All our experiments were done on a commodity laptop with 16 GB
RAM and an 8-Core Intel 2.3GHz CPU running macOS.

\vspace{-5pt}

\subsection{RQ1: Comparison to the State of the Art}

Multiple tools~\cite{JinSZLL11, Jin0D12, Liu-al:FSE14, alphaFixer, LinWLSZW18}
have been proposed to repair concurrency bugs.
Of those, \pfix~\cite{LinWLSZW18} is the most recent concurrency bug-fixing tool
for Java programs, shown to significantly outperform the previous
state-of-the-art, namely \grail~\cite{Liu-al:FSE14}, in terms of efficiency,
correctness and patch quality.
Moreover, similar to our approach, \pfix also targets data races (along with
 atomicity violations). 
Therefore, we focus on comparing \tool with \pfix.

\paragraph*{Selection of Subject Applications}

We chose 24 unique subject apps from the benchmark suites universally
adopted by the APR community as a baseline for evaluating  tools which fix concurrency bugs~\cite{pecan,
  Jacontebe, fixexamples}.
Our choice of subjects is dictated by the following two aims: (a) to
evaluate \tool with regard to various data race patterns and (b) to
include all subjects from the \pfix suite, thus comparing to it on its
home turf.

Table~\ref{tab:tab_rq1} contains essential information about the
subject apps and provides the evaluation results of \pfix and \tool on
them.
In the following, we discuss the comparison of the two tools
\wrt the following four aspects:
\begin{enumerate}[label=(\alph*),topsep=0pt]

\item Bug detection efficiency: how many bugs were found.

\item Performance of both detecting and fixing bugs.

\item Quality of the produced concurrent patches.

\item Fundamental limitations of both approaches.

\end{enumerate}


\paragraph{Comparing bug detection efficiency}

\begin{figure*}[t]
\begin{subfigure}{0.95\textwidth}
  \centering
  \includegraphics[width=\linewidth]{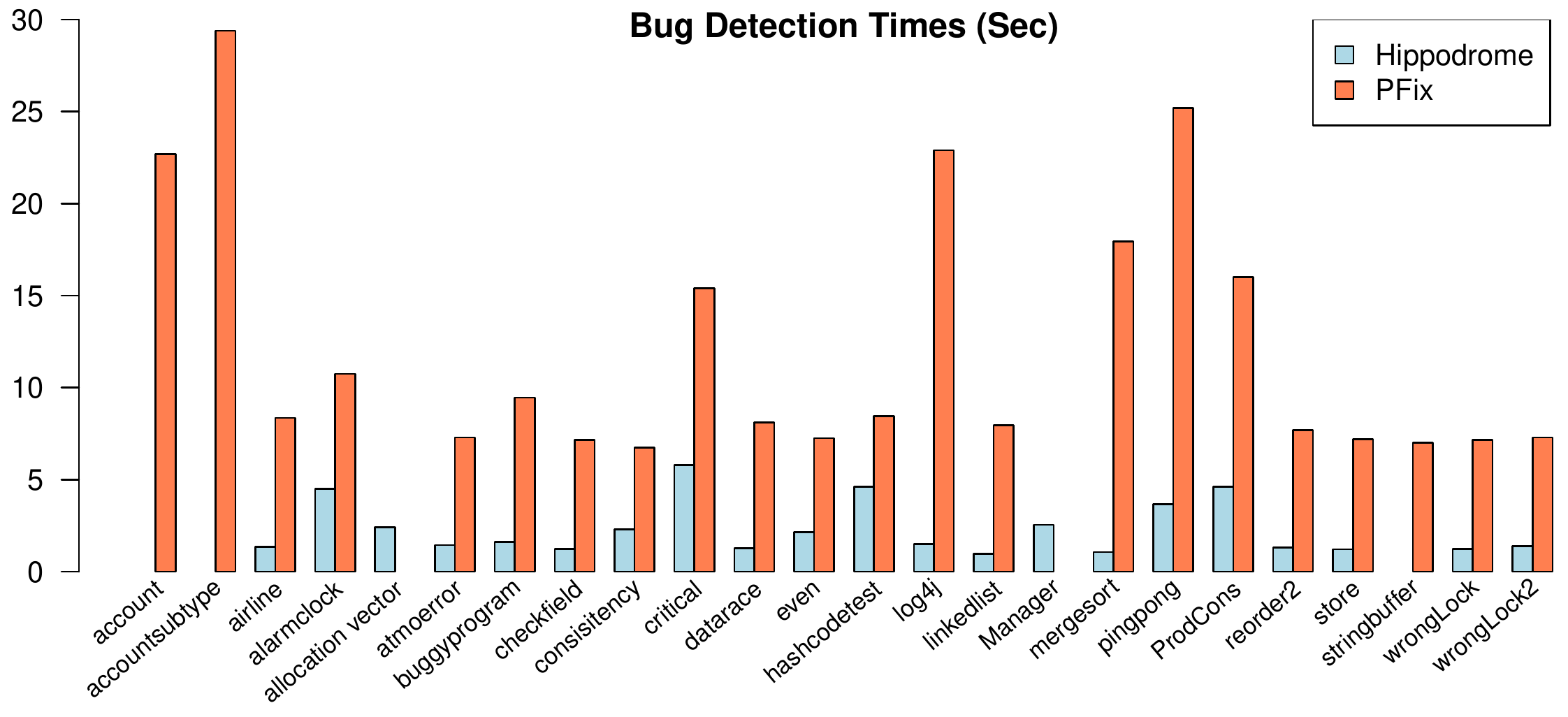}
  \setlength{\abovecaptionskip}{0pt}
  \caption{Bug detection times}
  \label{fig:sfig1}
\end{subfigure}%
\vfill
\begin{subfigure}{0.95\textwidth}
  \centering
  \includegraphics[width=\linewidth]{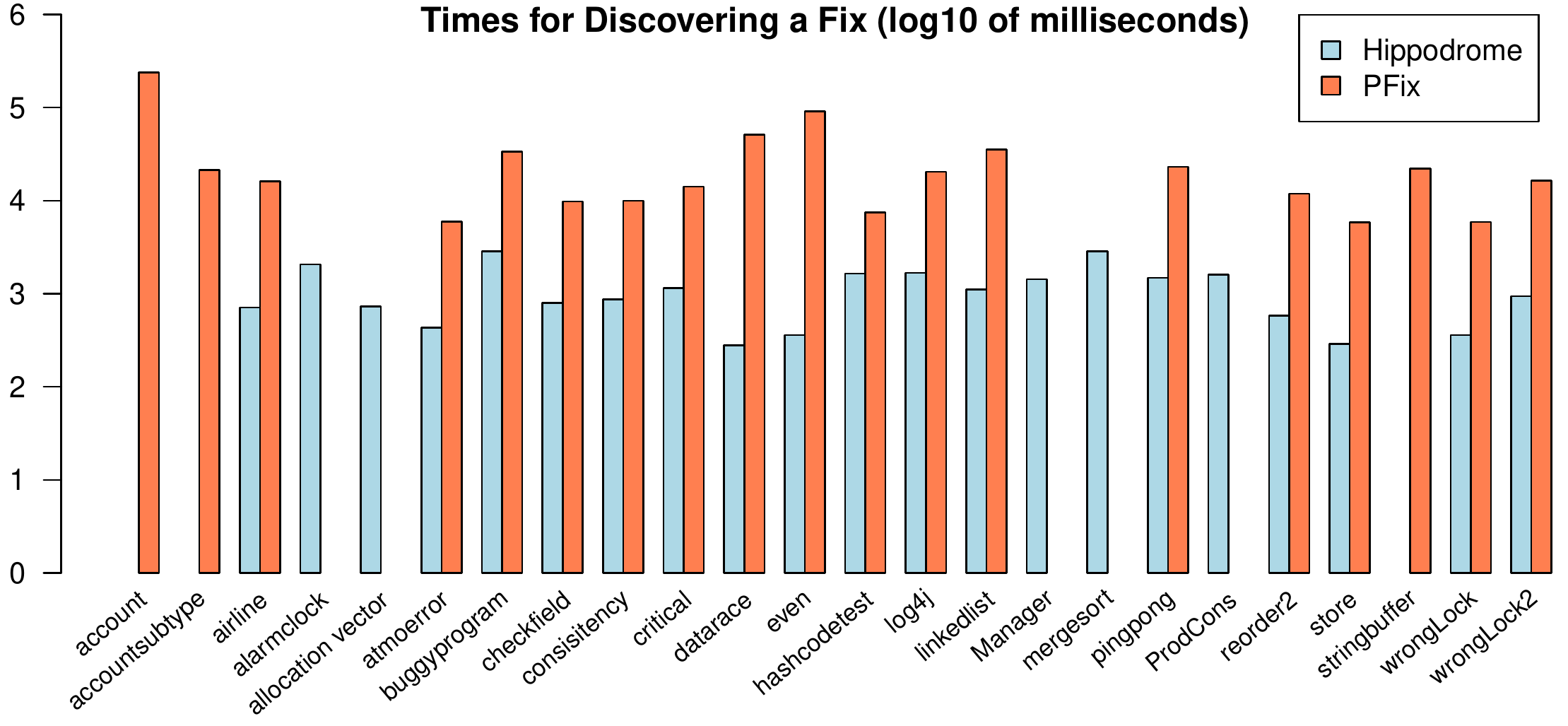}
  \setlength{\abovecaptionskip}{0pt}
  \caption{Time to find a fix (logarithmic scale)}
  \label{fig:sfig2}
\end{subfigure}
\setlength{\belowcaptionskip}{-10pt}
\setlength{\abovecaptionskip}{5pt}
\caption{\tool \vs \pfix: run-time comparison.}
\label{fig:runtimesub}
\end{figure*}

\tool leverages the enhanced version of \racerd
(\cf~\autoref{sec:analysis}) for bug detection and creates
repairs for \emph{all} detected bugs  
in the
selected subject apps.
The seventh column in Table~\ref{tab:tab_rq1} shows the 
detection rate on the subject programs.
In particular, \tool failed to fix three programs, all due to missed bugs. 
A close inspection 
revealed that all bug misses correspond to {atomicity violations} inside an
already synchronised block, 
which is consistent with the goal of only detecting
races.
%
 \racerd detected one of the two atomicity violations
~\cite{LuPSZ08} in \code{buggyprogram} (manifested as a race); thus,
\tool only fixed the detected violation---we mark this as a partial
fix.

To fix a data race, \pfix requires memory access patterns of passing
and failing test executions obtained by running
%
Java PathFinder~\cite{VisserPK04} and
Unicorn~\cite{unicorn} (both 100 times).
The patterns are ranked based on their occurrence in the test executions. Each run of \pfix generates a repair to be validated using 100 test runs. If the bug is still observed, PFix reruns continuing the process until a failure is not observed for all 100 test runs. 
\pfix failed to fix \emph{five programs} in our selection; in two
cases, JPF failed to obtain a failing test execution, and in three
instances, \pfix was unable to create a successful fix. 


\paragraph{Comparing runtime performance}
Figure~\ref{fig:runtimesub} compares the run-time performance of \tool to
\pfix. Subfigure~\ref{fig:sfig1} provides the detection time and
Subfigure~\ref{fig:sfig2} provides the time taken to fix. 
The total time (detection and fix) in \pfix does \emph{not} include the
time to validate a fix, whereas the total time of \tool also 
includes validation. 
\tool first detects all the bugs in a program, a process
which includes both the compilation and the
analysis.  
In 11 instances, \racerd was invoked twice, \ie, once to detect the
bugs and another to validate the fix. 
In other cases, it was invoked multiple times (within a single run of \tool) 
to validate different repair options. For example, in \code{critical} program, \tool inferred five
 patches, thus, \racerd ran six times and took 5.8
seconds.
As per \autoref{fig:sfig2}, \tool outperforms \pfix \emph{by
  several orders of magnitude}. 
  
\vspace{-5pt}

\paragraph{Comparing patch quality}

In 14 out of 24 cases, \tool produced repairs equivalent to \pfix's
patches. 
In four cases, \tool failed to detect bugs or succeed partially, while
in six it created better fixes than \pfix: the latter either
missed those bugs or created an incomplete fix. 
For example, \tool's repair fixes the data races in both
\code{withdrawal} and \code{deposit} methods in \autoref{fig:datarace}
by wrapping them into \code{synchronized (this)\ \{...\}}. 
The patch produced by \pfix ignores the data race in \code{withdrawal}  because its input
memory access patterns do not contain the bug path for
that method.
This example highlights the basic problem with tools relying on
dynamic approaches: identifying a subset of buggy paths might be
insufficient for producing a complete fix.

\vspace{-5pt}
\paragraph{Inherent downsides of \pfix and \tool}
Apart from the inherited inefficiencies from JPF (\eg, scalability to large Java
programs), \pfix's methodology suffers from randomness in the ranking of
failure-inducing memory access patterns. This results in several incorrect
repairs involving thus multiple validations and degrading efficiency. 
Besides, \pfix chooses a locking policy dynamically by monitoring
the lock acquisition patterns during the shared variable access. The
incomplete nature of this policy makes the suggested repairs 
 deadlock-prone (confirmed in~\cite[\S{4.4}]{LinWLSZW18}).
 On the other hand, \tool's expressivity is limited by the power of the
 underlying analysis, 
 which makes it applicable exclusively to repairing data races, but not, \eg,
 atomicity violations.

We conclude this case study by responding to \textbf{RQ1}:

\vspace{5pt}

\noindent\framebox{\parbox{\dimexpr\linewidth-2\fboxsep-2\fboxrule}{%
    \itshape \, \tool generates high-quality repairs for 20 out of 24
    reported bugs in a benchmark suite of a state-of-the-art tool
    \pfix. In all successful cases, \tool run-time significantly
    outperforms \pfix, while its patch quality is the same or better
    in the majority of the cases.}}
\vspace{3pt}

\subsection{RQ2: \tool and Human Fixes}

We evaluated \tool's performance, correctness and quality
of its patches on known concurrency bugs in large-scale open-source
Java projects. We first searched for the occurrences of the
keywords \emph{concur}, \emph{thread}, \emph{sync}, \emph{lock}, and
\emph{race} in the commit history of Apache Tomcat and Google Guava over the past five years.
Next, we ran \racerd on each version preceding the commit that fixed a
 bug
 and manually checked whether the fixed bug was detected.\footnote{Few cases
   required a single \texttt{@ThreadSafe} annotation to the buggy class.}
In particular, we analysed 50 concurrency bugs and
successfully reported 39 of those. The following table summarises our
findings:
\begin{center}
	{   
			\begin{tabular}{lcc}
				\toprule
				\textbf{Project Name} & \textbf{\#~Confirmed Bugs} & \textbf{\#~Detected by \racerd}
				\\ 
				\midrule
				Apache Tomcat         & 43                              & 35                           \\ 
				Google Guava              & 7                               & 4                            \\ 
				\bottomrule
			\end{tabular}  
	}
\end{center}

\paragraph{Statistics of developers' fixes}
\begin{table}[t]
	\caption{Syntax of the developers' fixes}
	\label{fig:devfixes}
	\begin{tabular}{lcccc}
		\toprule
		\textbf{Project} & \textbf{Added Volatile} & \textbf{Added sync block} & \textbf{Changed Collection} & \textbf{Others} \\
		\midrule
		Tomcat           & 32                      & 14                                & 7                                & 7               \\ 
		Guava            & 3                       & 4                                 & 2                                & 0               \\ 
		\bottomrule
	\end{tabular}
\end{table}

Table~\ref{fig:devfixes} summarises different kinds of fixes
introduced by developers.
In particular, 69 code changes were performed to repair 50 races.
%
%
Making a shared variable \code{volatile} was the popular approach (35
out of 69) to data race repair,
although they could have also been fixed via
\code{synchronized}.
%
%
At the same time, the developer's input is also needed to validate
the addition of \code{volatile}, as affected writes may become
expensive.
Because of this, \tool suggests adding \code{volatile} as a secondary
fix option, favouring the use of \code{synchronized} as a primary fix.
In few cases, developers changed the type of Java collections to
their thread-safe variants, \eg, \code{HashMap} to \code{ConcurrentHashMap}.
This last kind of fix is beyond the scope of this work.

\paragraph{
	Repair quality}
\begin{table}[]
	\caption{\tool results on Developers' commits.}
	\label{tab:rq2}
	\begin{adjustbox}{width=0.6\linewidth}
	\begin{tabular}{llllll}
		\cline{1-6}
		\textbf{Index} & \multicolumn{1}{|c|}{\textbf{Tomcat}}     & \multicolumn{4}{c}{\textbf{\tool}}                                                                       \\
		& \multicolumn{1}{|l|}{Commit\#}   & \multicolumn{1}{l|}{Fix Status} & \multicolumn{1}{l|}{Detection Time} & \multicolumn{1}{l|}{Fix Time} & Equivalence \\ \cline{1-6} 
		1     & 752f0b9f                        & Fail                            & NA                             & NA                            & NA          \\
		2     & 582cc729                        & Success                         & 87.92                          & 21.47                         & Syntactic   \\
		3     & 5e9f6fd6                        & Success                         & 88.25                          & 15.95                         & Syntactic   \\
		4     & 066e2546                        & Fail                            & NA                             & NA                            & NA          \\
		5     & 317480b9                        & Success                         & 82.29                          & 17.85                         & Semantic    \\
		6     & 7040497f                        & Success                         & 76.84                          & 19.32                         & Semantic    \\
		7     & 29f060ad                        & Fail                            & NA                             & NA                            & NA          \\
		8     & b96f9bec                        & Success                         & 77.94                          & 30.75                         & Semantic    \\
		9     & 0ca05961                        & Success                         & 14.75                          & 6.6                           & Syntactic   \\
		10    & be19e9b1                        & Fail                            & NA                             & NA                            & NA          \\
		11    & 4caec93b                        & Fail                            & NA                             & NA                            & NA          \\
		12    & ed610381                        & Success                         & 73.84                          & 23.28                         & Syntactic   \\
		13    & 057de944                        & Success                         & 71.09                          & 20.6                          & Semantic    \\
		14    & 518c27c3                        & Success                         & 72.3                           & 14.65                         & Semantic    \\
		15    & 227b6093                        & Success                         & 75.09                          & 7.84                          & Syntactic   \\
		16    & fb631d21                        & Success                         & 74.63                          & 6.34                          & Semantic    \\
		17    & 50121380                        & Success                         & 80.46                          & 17.57                         & Semantic    \\
		18    & d85c35f                         & Success                         & 77.29                          & 54.6                          & Semantic    \\
		19    & 3360c3a                         & Success                         & 56.28                          & 42.3                          & Semantic    \\
		20    & 8cbb9f8                         & Success                         & 112.61                         & 10.37                         & Semantic    \\
		21    & afff25f1c                       & Success                         & 55.84                          & 15.18                         & Syntactic   \\
		22    & d4c8da6                         & Success                         & 55.28                          & 20.67                         & Syntactic   \\
		23    & d9b530c                         & Success                         & 55.19                          & 19.15                         & Syntactic   \\
		24    & e80797f3                        & Success                         & 55.34                          & 18.67                         & Syntactic   \\
		25    & 1484f3ec                        & Success                         & 57.49                          & 7.55                          & Semantic    \\
		26    & 825c450c                        & Success                         & 55.27                          & 9.38                          & Semantic    \\
		27    & 02a4bb92                        & Success                         & 56.99                          & 4.13                          & Semantic    \\
		28    & 3d6dbd91                        & Success                         & 55.36                          & 15.36                         & Syntactic   \\
		29    & 8313fa0f1                       & Success                         & 52.57                          & 17.4                          & Syntactic   \\
		30    & 52b29fd2                        & Success                         & 54.82                          & 21.3                          & Syntactic   \\
		31    & ea383db5                        & Success                         & 55.39                          & 11.46                         & Syntactic   \\
		32    & cadbc500                        & Fail                            & NA                             & NA                            & NA          \\
		33    & 5379ae68                        & Success                         & 55.4                           & 19.67                         & Syntactic   \\
		34    & 3078444                         & Success                         & 52.27                          & 9.86                          & Syntactic   \\
		35    & 17b6c64f                        & Success                         & 55.17                          & 15.48                         & Syntactic   \\ \cline{1-6} 
		& \textbf{Guava} &                                 &                                &                               &             \\ \cline{1-6} 
		36    & 0e94fb5b                        & Success                         & 24.74                          & 10.033                        & syntactic   \\
		37    & 0e94fb5b                        & Success                         & 24.67                          & 80.02                         & syntactic   \\
		38    & c15cd804                        & Fail                            & NA                             & NA                            & NA          \\
		39    & a43b4aa7                        & Fail                            & NA                             & NA                            & NA          \\ \cline{2-6} 
	\end{tabular}
\end{adjustbox}
\end{table}
Table~\ref{tab:rq2} provides the \tool's results on 39 developers' commit. Column \emph{Commit\#} describes the commit under which the developer fixed a particular concurrency bug. \emph{Fix status} denotes whether \tool successfully fixed a bug, and \emph{Detection/Fix Time} shows the time taken to detect/fix a bug. Finally, \emph{equivalence} shows the syntactic and semantic equivalence of the repair.
\racerd failed to detect 8 bugs in Tomcat (out of 43) and 3
 in Guava (out of 7); thus, we ran \tool on 39 remaining bugs
(detailed statistics 
in the accompanying supplementary
material).
\tool generated 83\% correct fixes for Tomcat (29/35), and 50\% for
Guava (2/4). 
In particular, it failed to repair 3 bugs requiring to add
interface-level annotations which 
are beyond the
scope of this work.
In the remaining 5 fixes, developers introduced new features, 
thus we cannot draw a  direct 
comparison to those.

\begin{figure}[t]
  \includegraphics[width=1.1\textwidth]{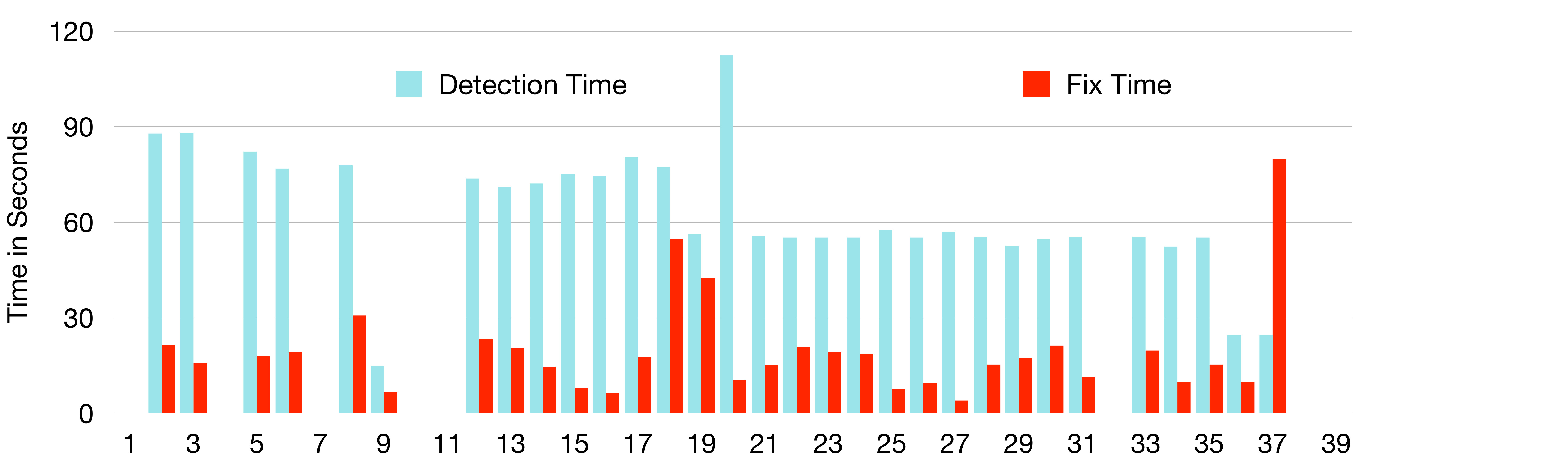}
  \caption{\tool run-time on large-scale projects}
  \label{fig:runtimerq2}%
\end{figure}

\autoref{fig:runtimerq2} shows \tool's performance on the selected
39 bugs. 
Detection and validation took about one minute,
most of which is attributed to
compilation.\footnote{In its standard mode, \racerd attaches itself to
	the compilation process.}
\racerd detection time is usually shared to detect multiple bugs, yet in this
experiment we ran it from a ``cold'' setup, zooming in on each particular bug.
The AST creation for project files takes most of the time to find a fix;
%
the actual fix generation takes less than a
second.
Therefore, amortised times for detecting and fixing several bugs in a
batch mode would be much smaller.

Repairs for 18 bugs were syntactically the same as the developers'
fixes.
In 2 cases \tool generated better repairs, introducing \emph{less} overhead.
In one instance,  the developer's repair was better: our repair locked multiple
\code{case} clauses in a \code{switch} statement, instead of just the race-causing clause.
%

An attempt to run \pfix on these large-scale projects produced no fix report irrespective of the set timeout.\footnote{This is due to inherent scalability limitations of JPF, which we explicitly confirmed in our communication with the \pfix authors.}

\vspace{5pt}

\noindent\framebox{\parbox{\dimexpr\linewidth-2\fboxsep-2\fboxrule}{%
    \itshape \, \tool repairs 31 out of 39 concurrency bugs. 60\% of
    these fixes are syntactically the same as the developer patches.
    \tool runtime performance scales well for large-projects with about 80 seconds to fix the first bug.}}

\section{Threats to Validity}
\label{sec:disc-limit}

\paragraph{Completeness}

Completeness of our approach (\ie, whether it only repairs \emph{real} data
races) relies on the underlying analysis. \racerd is proven to report \emph{no
  false positives} under the concurrency model restricted to balanced
reentrant locks and coarse-grained lockings~\cite{GorogiannisOS19}. 
This model has been shown 
to provide high signal in production
code~\cite{BlackshearGOS18}.

\paragraph{Soundness}
The soundness of our approach (\ie, whether it catches \emph{all} data 
races and fixes them without introducing new issues) relies on the soundness
guarantees of \racerd and the soundness of our patch inference algorithm.
The assumptions under which the analysis does not miss
races are listed in~\cite[{\S}8]{BlackshearGOS18}. Specifically, it assumes that
an owned access and its suffixes remain owned throughout
 the current procedure. Violating this assumption can lead to false
negatives if a local object is leaked to another thread \emph{and} if the other
thread accesses one of the objects fields.

\tool's patch inference 
relies on its validation
pha-se to avoid deadlocks. Patching is sound in that it introduces no
deadlocks with two crucial, yet reasonable, assumptions:
\begin{enumerate}[label=(\alph*)]
\item Locking is \emph{balanced}, \ie, lock releases follow a LIFO order. This
  effectively corresponds to the case of Java's \code{synchronized}, which
  only allows lexically-scoped locks.
\item Lock-acquiring code is located within \emph{non-anonymous} classes, since
  summaries have to be ascribed to methods of named classes. Therefore, deadlocks
  may be missed when the locking code is inlined in the spawned
  threads.
\end{enumerate}
Assumption (b) is satisfied by industrial software, which always implements
synchronisation \emph{in libraries}, while spawning threads in thread pools.
However, micro-benchmarks used to evaluate dynamic
analysis-based tools might violate it as they favour concise and self-contained code 
for tractable dynamic race detection.
At the time of this submission, the static analysis team at Facebook confirmed
that both race and deadlock detection were running in production for more than
two years \emph{with no false negatives
identified}~\cite{Brotherston20}.\footnote{The deadlock analysis
is  proven sound for a  language with scoped re-entrant locks,
nondeterministic branching, and non-recursive procedure calls.}
%


\section{Related Work}
\label{sec:related}

\paragraph*{General Program Repair and Synthesis}
Automated program repair (APR)~\cite{GouesPR19} is an emerging technology paradigm for automatically fixing logical bugs via search \cite{GenProg09}, semantic reasoning \cite{Semfix13} and learning \cite{Prophet16}. The recent works on \emph{semantic} program repair~\cite{Semfix13,Angelix16} make use of advances in program synthesis \cite{PolikarpovaS19,Sergey18, CosteaZPS20,NguyenTSC21} to automatically generate one-line or multi-line fixes.  However, these approaches have been mostly studied for sequential programs.


\paragraph*{Program Repair for Concurrency}

 While \pfix \cite{LinWLSZW18} is a recent work in this direction, there are other efforts on concurrency repair, such as~\cite{alphaFixer,ConTest,ConTest2,JinSZLL11,Jin0D12,Liu-al:FSE14,LiuCL16,LiuZ12,LinWLSZW18,HuangZ12a} 
 to name a few.  
 Many of these tools use dynamic analysis to find bugs, while some statically validate fixes~\cite{JinSZLL11,Jin0D12}. 
 In general, dynamic analysis approaches may miss concurrency bugs, and the repairs may be incomplete.
  \pfix \cite{LinWLSZW18} denotes a somewhat hybrid approach which relies on the JPF model checker for bug finding. 
  Apart from the scalability limitations of JPF, such an approach requires providing the temporal properties to the model checker---which \pfix provides by exploiting an incomplete set of likely failure inducing memory access patterns. 
  The work on \hfix~\cite{LiuCL16} falls into the category of using syntactic or pattern-matching based static analysis, where patterns of patches are obtained by mining human patches. 
  Such an approach is inherently limited to the data-set of human patches considered, and carry no guarantees of correctness.
   \grail, also based on static analysis~\cite{Liu-al:FSE14}, offers patches with deadlock-free guarantees,
   although these guarantees have been shown by \cite{LinWLSZW18} not to always hold in practice.
   Unlike our work, \grail is not modular; it relies on mixed-integer programming computation and Petri net models which restrict the fixes to single classes or methods. 
   Such bugs which span multiple classes or methods pose no problem for Hippodrome as shown by our experiments.

\paragraph*{Program Repair with Static Analysis}
\label{sec:program-repair-with}

Logozzo and Ball co-authored an early work using abstract interpretation based static analysis for sequential program repair~\cite{Logozzo}. Unlike our work which requires no user annotations, their approach needs formal safety properties to drive the repair.
The work on \afix focuses on atomicity violations~\cite{JinSZLL11}. \afix, along with its later incarnation \tname{CFix}~\cite{Jin0D12}, represent pattern-based approaches for fixing concurrency bugs; these techniques do not come with correctness guarantees. \afix is in fact known to introduce deadlocks as pointed out in \cite{Liu-al:FSE14}.  Similarly, the work on \tname{Phoenix}~\cite{phoenix19} also patches static analysis warnings via repair strategy examples learnt from big codebases. The recent work on \footpatch is closer to our theme of analysis-driven correct repairs~\cite{TonderG18}. \footpatch exploits the \emph{locality} in the summaries inferred by a compositional static analysis~\cite{CalcagnoDOY09} for efficiently generating sound one-line patches for imperative programs, fixing memory leaks and null-pointer exceptions. Due to a non-local nature of concurrency issues (\eg, data races), \footpatch cannot be used directly for fixing concurrency errors, which is achieved in this paper.

\vspace{-3pt} 

\section{Conclusion}
\label{sec:conclusion}

We have described a static analysis driven APR technique for concurrent programs
which is scalable, modular, and preserves deadlock-freedom. It can be fitted
into a Continuous Integration (CI) loop, which can allow collaboration with
developers to gradually improve automatically generated patches. 

Our publicly available tool, \tool, can be used to further explore related research avenues: richer solution space with more synchronization primitives, application-dependent cost functions for patches, lock choice  heuristics. It can possibly also be used to create different user studies in order to understand solution comprehension and preference.

\bibliographystyle{ACM-Reference-Format}
\bibliography{references}

\end{document}